\documentclass{aa}
\usepackage{graphicx}
\usepackage{longtable}
\usepackage{txfonts}
\begin{document}
\title{The embedded clusters DBS\,77, 78, 102, and 160$-$161 and their link with the interstellar medium\thanks{Based on observations gathered as part of observing programs: 179.B-2002,VIRCAM, VISTA at ESO, Paranal Observatory; 087.D-0490A, NTT at ESO, La Silla Observatory and CN2012A-045, SOAR telescope at NOAO, CTIO.}}
\author{M.A. Corti\inst{1,2}, G.L. Baume\inst{1,3},  J.A. Panei\inst{1,3}, L.A. Suad\inst{2}, J.C. Testori\inst{2}, J. Borissova\inst{4,5}, R. Kurtev\inst{4,5}, A.N. Chen\'e\inst{6} \& S. Ramirez Alegr\'{\i}a\inst{4,5}}
\authorrunning{Corti et al. }
\titlerunning{Embedded clusters and the ISM}
\institute{Facultad de Ciencias Astron\'omicas y Geof\'{\i}sicas,
Universidad Nacional de La Plata, Paseo del Bosque s/n, B1900FWA La Plata, Argentina
\and 
Instituto Argentino de Radioastronom\'{\i}a (CCT-La Plata, CONICET), C.C. No. 5, 1894 Villa Elisa, Argentina
\and 
Instituto de Astrof\'{\i}sica de La Plata (CCT-La Plata, CONICET - UNLP), Paseo del Bosque s/n, B1900FWA La Plata, Argentina
\and
Instituto de F\'isica y Astronom\'ia, Universidad de Valpara\'iso, Av. Gran Breta\~na 1111, Playa Ancha, Casilla 5030, Chile
\and
Millennium Institute of Astrophysics (MAS), Av. Gran Breta\~na 1111, Playa Ancha, Casilla 5030, Chile
\and
Gemini Observatory, Northern Operations Centre, 670 North A'ohoku Place, Hilo, HI 96720, USA}
   \date{Received ***; accepted ****}
  \abstract
   {}
{We report a study of the global properties of some embedded clusters placed in the fourth quadrant of the Milky Way
 to clarify some issues related with their location into de Galaxy and their stellar formation processes.}
{We performed $BVI$ photometric observations in the region of {\rm DBS}\,77, 78, 102, 160, and 161 clusters and infrared spectroscopy in {\rm DBS}\,77 region. They were complemented with $JHK$ data from {\rm VVV} survey combined with {\rm 2MASS} catalogue, and used mid-infrared information from {\rm GLIMPSE} catalogue. We also searched for H\,I data from {\rm SGPS} and {\rm PMN} radio surveys, and previous spectroscopic stellar classification. The spectroscopic and photometric information allowed us to estimate the spectral classification of the brightest stars of each studied region. 
 On the other hand, we used the radio data to investigate the interstellar material parameters and the continuum sources probably associated with the respective stellar components.}
 {We estimated the basic physical parameters of the clusters (reddening, distance, age, and initial mass function). We searched for HII regions located near to the studied clusters and we analyzed the possible link between them. In the particular case of {\rm DBS}\,160$-$161 clusters, we identified the H\,I bubble {\rm B}332.5-0.1-42 located around them. We found that the mechanical energy injected to the interstellar medium by the more massive stars of this couple of clusters was enough to generate the bubble.}
 {} 
   \maketitle
%
\section{Introduction} 
\label{sec:intro}
Embedded clusters provide an important tool to investigate stellar properties, the interstellar medium (\rm {ISM}), and the structure of the Galaxy \citep{pinh12}. In particular, several embedded clusters include massive stars and present important pre main sequence (PMS) populations \citep{che13}. They allow us to study early$-$type stars, their impact on the surrounding environment and help to outline the path of the spiral arms. The O and B type stars of these clusters emit large amounts of energy which modify the properties of the surrounding ISM. These interactions cause photodissociation regions like H\,II regions \citep{leq05}. The winds of the early stars can also originate bubbles, identified as a minimum in the H\,I emission distribution surrounded by regions of higher emissivity. These H\,I structures and H\,II regions are signatures of star formation. 

Since embedded clusters are immersed in a large amount of gas and dust, they are affected by high visual extinction. This makes necessary to employ infrared (IR) observations for a proper study. However, optical data, at least for the brightest cluster members, are an important tool to obtain reliable values of the corresponding color excess, absorptions and reddening behavior. 

During the last years, several systematic searches of embedded clusters have been developed based on new infrared sky surveys. In particular, those works using Vista Variables in the V\'{\i}a L\'actea ({\rm VVV})\footnote{http://www.vista.ac.uk} survey \citep[e.g.][]{bor11,bor14,barb15}. Our goal is to obtain the characteristic parameters of a large sample of embedded clusters using homogeneous data and analysis methods. Results for four 
known clusters were presented by Chen\'e et al. (2012) and for six of the new {\rm VVV} cluster candidates with Wolf-Rayet {\rm (WR)} stars in Chen\'e et al. (2013). In this paper we follow similar analysis methods as in those papers, but we applied them to a sample of clusters of the {\rm DBS} catalogue \citep{dut03} and we complemented them with optical and radio data.

In the present work we investigated the regions of the embedded clusters {\rm DBS}\,77, {\rm DBS}\,78, {\rm DBS}\,102, and {\rm DBS}\,160-161. In order to estimate the fundamental parameters of the clusters and their interaction with the \rm {ISM}, we performed a multifrequency study using optical, infrared, and radio data. All the selected regions are located in the galactic plane in the fourth quadrant of the Galaxy (see Table \ref{tab:param}). They are probably associated with identified H\,II regions and some particular objects. In the following we present a brief description of them:

\begin{itemize}

\item Embedded cluster {\rm DBS}\,77 is placed near the recently identified clusters {\rm VVV}\,15 and {\rm VVV}\,16 \citep{bor12} using {\rm VVV}  survey. These three clusters are located close to the dark nebula Dobashi 5898 \citep{dob11} that was revealed from near infrared observations from {\rm 2MASS} data.
They seem to be associated with the IRAS source 12320-6122 and placed inside the H\,II region {\rm RCW}\,65 = Gum\,43. All the complex is inside the molecular cloud G301.0+1.2. They are almost in the same line of sight that the foreground open cluster Ruprecht\,105 placed at 950 pc \citep{kha05}. 

\item Embedded cluster {\rm DBS}\,78 is associated with the IRAS source 12331-6134 and placed inside the ultra compact H\,II region {\rm GRS}\,G301.11+00.97 \citep{bro96}. They are located in the molecular cloud G301.1+1.0 \citep{rus04} and close to the dark nebula Dobashi 5905. 

\item Embedded cluster {\rm DBS}\,102 is in the core of the H\,II region {\rm G}333.0+0.8 \citep{kuc97} which is surrounded by several ($\sim$ 20) dark and molecular clouds identified by the analysis of mid infrared images produced by Spitzer, available via Galactic Legacy Infrared Midplane Survey Extraordinaire ({\rm GLIMPSE})\footnote{http://www.astro.wisc.edu/glimpse/}. The analysis of the Multiband Infrared Photometer for Spitzer Galactic survey ({\rm MIPSGAL})\footnote{http://irsa.ipac.caltech.edu/data/SPITZER/MIPSGAL}  data \citep{per09} is another information source. These molecular clouds were also identified by the CO observations \citep{rus04}. Close to the cluster is located the dark nebula Dobashi 6419. In all these clouds young stellar objects (YSOs) have been identified from {\rm GLIMPSE} data \citep{rob08}. 

\item Embedded clusters {\rm DBS}\,160 and {\rm DBS}\,161 form a couple located at the center of the H\,II region G332.5$-$00.1 \citep{kuc97, bro96} and are identified with the IRAS source 16132-5039 which is also surrounded by about 20-30 dark and molecular clouds identified by CO observations \citep{rus04}. The analysis of mid infrared images produced by Spitzer and {\rm MIPSGAL} data is another information source. There are also three sources with IR excess according to the Midcourse Space Experiment ({\rm MSX}) data \citep{ega03}. All this region is part of the extended star formation complex {\rm RCW} 106 \citep{bai06}. These two clusters have been studied by \citet{rom04} using near infrared (NIR) photometric observations and by \citet{rom09} through the spectral classification of three stars in the region. Two of them identified as early main sequence (MS) stars and the third one classified as a YSO. 

\end{itemize}

This paper is organized as follows: Section \ref{sec:data} reports the different sources of the analyzed data and the way they were processed. Section \ref{sec:analysis} describes the multifrequency study carried out over the described data. Section \ref{sec:clustersregions} presents a description of the characteristics found on each cluster region. In Section \ref{sec:discussion} we discuss the interpretation of some particular results and in Section \ref{sec:conclusions} we present our first conclusions.
\begin{figure*}[!ht]
\centering
\includegraphics[width=12cm]{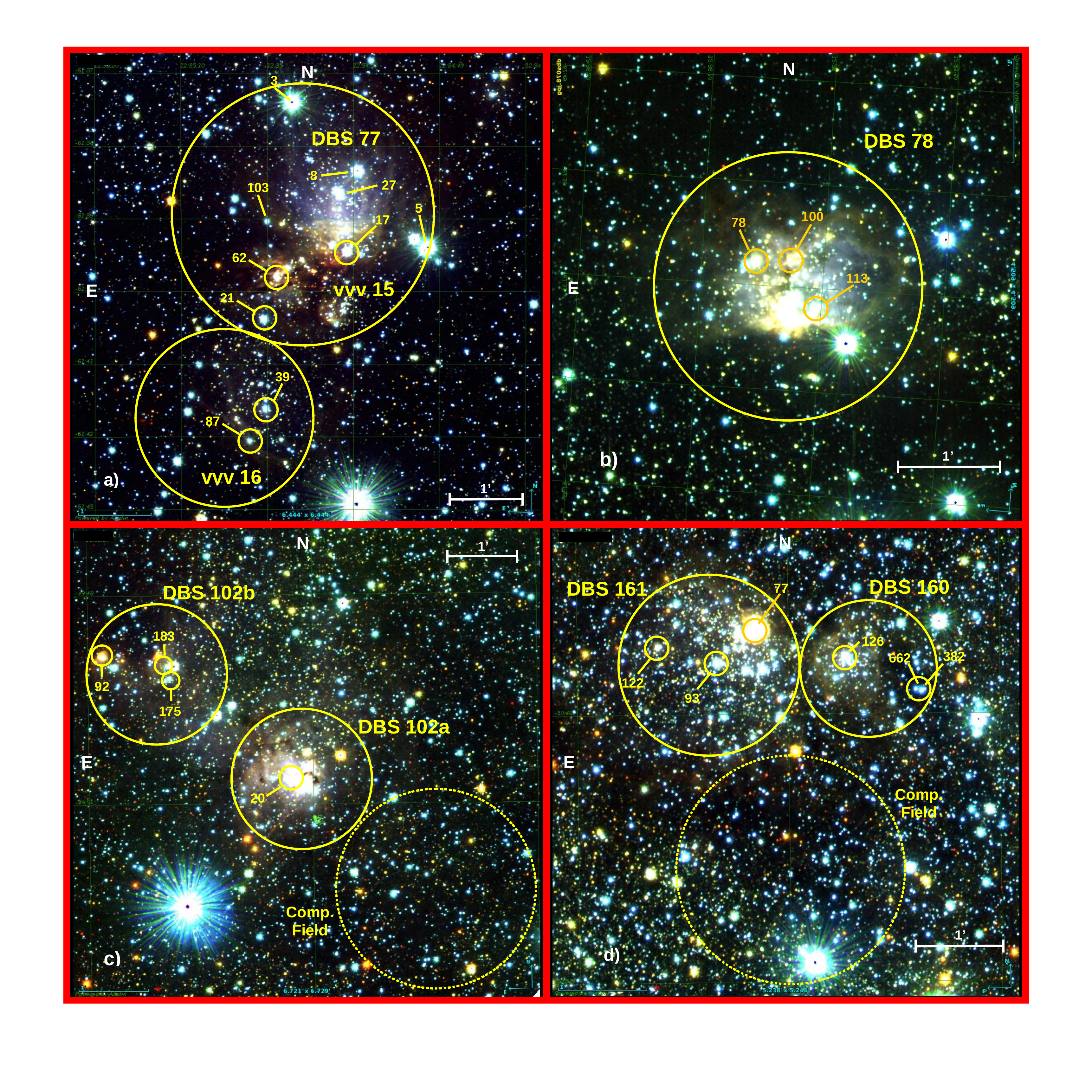}
\caption{$JHK$ false-color VVV images of the four studied regions. Selected clusters areas are indicated by big yellow solid circles. Comparison fields are also indicated with dotted circles in panels c) and d). Identified stars are those ones with NIR spectroscopy data or adopted brightest clusters members. Stars inside circles are adopted clusters members. (see text and Table \ref{tab:sptphot}).} 
\label{fig:charts}
\end{figure*}

\section{Data}
\label{sec:data}
This study makes use of the following material:
\begin{itemize}
\item $BVI_C$ images obtained by us with the Southern Astrophysical Research ({\rm SOAR})\footnote{http://ast.noao.edu/facilities/soar} telescope. 
\item Data from the following surveys/catalogues: a) 
The {\rm VVV} survey \citep{min10, sai12}, b) The {\rm APASS}\footnote{http://www.aavso.org/apass} catalogue \citep{hen10} of the American Association of Variable Star Observers ({\rm AAVSO}) c) The Two Micron All-Sky Survey ({\rm 2MASS})\footnote{http://www.ipac.caltech.edu/2mass/}, \citep{skr06}, d) The GLIMPSE catalogue from Spitzer Space Telescope (SST) data. e) Infrared spectroscopy gathered from New Technology Telescope ({\rm NTT})\footnote{https://www.eso.org/sci/facilities/lasilla/telescopes/ntt.html} at the European Southern Observatory ({\rm ESO}), La Silla Observatory and f) The Southern Galactic Plane Survey ({\rm SGPS})\footnote{http://www.atnf.csiro.au/research/HI/SGPS/queryForm.html} \citep{mcc05} and the Parkes-MIT-NRAO ({\rm PMN}).\footnote{http://www.parkes.atnf.csiro.au/observing/databases/PMN/PMN.html}
\end{itemize}

\begin{table*}
\fontsize{10} {14pt}\selectfont
\caption{Main parameters of the studied embedded clusters}
\centering
\begin{tabular}{ccccllllll}
\hline\hline
 Cluster & $\alpha_{J2000}$ & $\delta_{J2000}$   & $R$  & $V_0-M_V$ & $E_{B-V}$ & SpT$^{\ast}$ & U$^{\diamond}$ & Age$^{\circ}$ & $\Gamma^{\bullet}$ \\
 ID      & $[h:m:s]$        & $[^{\circ}:^{\arcmin}:^{\arcsec}]$ & $[\arcmin]$ & $[mag]$   & $[mag]$   &   & [pc cm$^{-2}$] & [Myr]   &  \\
\hline
\hline
 DBS~077  & 12:34:52.0 & -61:39:00.0 & 1.75 & 13.2 & 1.70 & O7 V$^{\triangle}$   & 240$^{\triangle}$ & $\sim 6.4^{\triangle}$ & $ -1.44 \pm 0.14^{\triangle}$ \\
 VVV~016  & 12:35:00.0 & -61:41:40.0 & 1.00 & 13.2 & 1.70 &                    &                   &                       &                              \\
\hline
 DBS~078  & 12:36:03.0 & -61:51:00.0 & 1.30 & 13.2 & 3.00 & B1.4 V &  7 &  $<$ 20 & $ -1.19 \pm 0.22$ \\
 \hline
 DBS~102a & 16:15:01.0 & -49:50:41.0 & 1.00 & 12.6 & 2.50 & O5$^{\triangle}$   & 78$^{\triangle}$ & $\sim 4^{\triangle}$   & $ -0.75 \pm 0.27^{\triangle}$ \\
 DBS~102b & 16:15:13.6 & -49:49:07.0 & 1.00 & 12.6 & 2.50 &     &    &   &      \\
 \hline
 DBS~160  & 16:16:55.5 & -50:47:26.0 & 0.75 & 12.3 & 2.50 & B0 V$^{\triangle}$   &  49$^{\triangle}$ & <10$^{\triangle}$     & $ -0.71 \pm 0.61^{\triangle}$ \\
 DBS~161  & 16:17:05.5 & -50:47:29.0 & 1.00 & 12.3 & 2.50 &                    &                   &                       &                              \\
 \hline
\label{tab:param}
\end{tabular}
\vspace{-0.6cm}
\tablefoot{ 
\tablefoottext{$^{\ast}$}{From the earliest MS star on each cluster region.}
\tablefoottext{$^{\triangle}$}{Parameter obtained considering both clusters as one.}
\tablefoottext{$^{\diamond}$}{Ionization parameter (see Eq. \ref{photons}, sec. \ref{sec:radiomaps})}
\tablefoottext{$^{\circ}$}{Upper age limit \citep{eks12}} 
\tablefoottext{$^{\bullet}$}{IMF slope obtained with the MS and PMS stars.}
}
\end{table*}

\subsection{Images}
\subsubsection{Optical data}
We used $BVI_C$ images that we acquired using the {\rm SOAR} Optical Imager (SOI)
mounted at the {\rm SOAR} 4.1m telescope at Cerro Tololo Inter$-$American Observatory ({\rm CTIO}, Chile). This camera has a mini-mosaic of two thinned and back illuminated E2V 2k $\times$ 4k CCDs.
We set a binning factor of 2 $\times$ 2 and then we obtained a scale value of $0\farcs153/pix$, and the 
mosaic covers, approximately, a field of view ({\rm FOV}) of $5\farcm2 \times 5\farcm2$. We obtained three shifted images on each field for each filter (see also Table~\ref{tab:frames1}). Images were acquired in 2012 during the night of May 15$^{\rm th}$. The typical full width 
at half maximum ({\rm FWHM}) was $\sim$ 2$^{\arcsec}$ and airmass values were about 1.3-1.5.

All frames were pre$-$processed in the standard way using the IRAF\footnote{IRAF is distributed by NOAO, which is operated by AURA under cooperative agreement with the NSF.}  task {\it ESOWFI/MSCRED}. That is, instrumental
effects were corrected with calibration images (bias and sky$-$flats taken during the same observing night). The exposures for each band were combined using {\rm IRAF} {\it MSCIMAGE} task. This procedure allowed us to fill the inter-chip gaps of the individual images. It also was useful  to remove cosmic rays and to improve the signal to noise ({\rm S/N}) for the final images.

\subsubsection{Infrared data}
In order to study the photometric behavior of the selected clusters (see Table~\ref{tab:param}), we used the stacked images of the individual $10\farcm0 \times 10\farcm0$ exposures containing the selected clusters from the VISTA Science Archive (VSA website\footnote{http://horus.roe.ac.uk/vsa/}; see \citealt{sai12} for more details about {\rm VVV} data).

\subsubsection{Astrometry}
World Coordinate System ({\rm WCS}) header information was available only for {\rm VVV} images, therefore we used $ALADIN$ tool and {\rm 2MASS} data to obtain this information for our optical images. Our adopted procedure to perform the astrometric calibration of our data was explained in \citet{bau09}. 
This allowed us to obtain a reliable astrometric calibration. The rms of the residuals in the positions were $\sim$ 0$\farcs16$, 
which is about the astrometric precision of the 2MASS catalogue ($\sim$ 0$\farcs12$).

\subsubsection{Photometry}
Instrumental magnitudes were obtained using IRAF {\it DAOPHOT} package.
We search for stars on the optical and infrared images using {\it DAOFIND} task. In the optical case, this task was performed over a white image obtained adding all the images for individual filters. We employed the point spread function (PSF) method \citep{ste87} on the $BVI$ and $JHK$ images. The PSF for each image was obtained from several isolated, spatially well distributed, bright stars ($\sim$ 10). The PSF photometry was aperture-corrected for each filter. Aperture corrections were computed performing aperture photometry of the same stars used as PSF models. All resulting tables were combined using {\it DAOMASTER} code \citep{ste92} obtaining one set for $BVI$ bands and other for $JHK$ ones.

Unfortunately, the night of the optical images was not photometric. Therefore our $BVI_C$ data were calibrated using the photometric values provided by {\rm APASS} catalogue, that is $B$ band and Sloan $gri$ bands for $\sim$10$-$20 stars on each cluster region. We transformed the $gri$ data to the $VI_C$ system using the equations provided by \citet{jes05}. To tie our observations to the standard system, we used transformation equations of the form: 
\begin{equation}
b = B + b_1 + b_2 (B-V)  
\end{equation}
\begin{equation}
 v = V + v_1 + v_2 (B-V)  
 \label{vmag}
\end{equation}
\begin{equation}
i = I + i_1 + i_2 (V-I)  
\end{equation}

\noindent where $BVI$ and $bvi$ are the transformed {\rm APASS} and instrumental magnitudes, respectively.

The calibration of infrared $JHK$ data was done using the information of common stars with {\rm 2MASS} catalogue on each cluster region. To join the obtained magnitudes from the {\rm VVV} images, we tried to use transformations with the form given by \citet{sot13}. However we noticed in some cases, a dependence of the residuals with the stellar magnitude. Therefore we adopted the following dependences:
\begin{equation}
j = j_1 + j_2 J + j_3 (J - H)
\end{equation}
\begin{equation}
h = h_1 + h_2 H + h_3 (J - H)
\end{equation}
\begin{equation}
k = k_1 + k_2 K + k_3 (J - K)
\end{equation}

\noindent where $JHK$ and $jhk$ are, respectively, {\rm 2MASS} and instrumental {\rm VVV} magnitudes.

The calibration coefficients of optical an infrared calibration equations were computed using $FITPARAMS$ task of {\rm IRAF} $PHOTCAL$ package. The obtained values are shown at Table~\ref{tab:frames2}. 

In order to complete our data for saturated stars in {\rm VVV} images, we simply adopted the {\rm 2MASS} magnitudes for the brightest objects ($K < 11$).

Regarding the errors of our photometry, for $BV$ and $JHK$ bands they can be estimated from their corresponding $rms$ values obtained in the fittings 
(see Table~\ref{tab:frames2}) and complemented with the error values provided by $DAOPHOT$ and $DAOMASTER$ codes for each particular object 
(see Sect.~\ref{sec:catalogues}). 
For the special case of $I$ band, these errors must be also complemented with the system transformation error ($\sim$0.03; \citealt{jes05}).

\begin{table}
\fontsize{10} {14pt}
\selectfont
\caption{Detail of optical and IR scientific frames}
\centering
\begin{tabular}{ccc}
\hline\hline
$Band$  & $exptime [sec]\times N^{\bullet}$ \\
\hline
$B$   & ~20 $\times$ 3 \\
$V$   & ~20 $\times$ 3 \\
$I_C$ & 100 $\times$ 3 \\
\hline
$J$   & ~10 \\
$H$   & ~10 \\
$K$   & ~10 \\
\hline
\label{tab:frames1}
\end{tabular}
\vspace{-0.6cm}
\tablefoot{ 
\tablefoottext{$^{\bullet}$}{$N =$ amount of observed frames}
}
\end{table}

\begin{table*}
\fontsize{10} {14pt}
\selectfont
\caption{Calibration coefficients used for optical and infrared observations}
\centering
\begin{tabular}{lrrrr}
\hline\hline
$Coef.$ & \multicolumn{1}{c}{DBS} & \multicolumn{1}{c}{DBS} & \multicolumn{1}{c}{DBS} & \multicolumn{1}{c}{DBS}     \\
        & \multicolumn{1}{c}{077} & \multicolumn{1}{c}{078} & \multicolumn{1}{c}{102} & \multicolumn{1}{c}{160-161} \\
\hline
$b_1$   &  0.48 $\pm$ 0.02 &  0.94 $\pm$ 0.01 & -0.23 $\pm$ 0.02 & --    \\
$b_2$   &  -0.13 $\pm$ 0.02 &  0.10 $\pm$ 0.02 &  -0.13 $\pm$ 0.02 & --               \\
$rms$   &  0.03  &  0.03  &  0.02   & --               \\ 
\hline
$v_1$   &  0.01 $\pm$ 0.02 &  1.52 $\pm$ 0.02 &  -0.25 $\pm$ 0.02 & --               \\
$v_2$   &  0.09 $\pm$ 0.02 &  0.29 $\pm$ 0.02 & 0.07 $\pm$ 0.02 & --               \\
$rms$   & 0.04    & 0.04    &  0.02        & --               \\
\hline
$i_1$   &  0.80 $\pm$ 0.05 & 1.10 $\pm$ 0.03 &  --              & --               \\
$i_2$   &  0.01 $\pm$ 0.04 &  0.17 $\pm$ 0.02 &  --              & --               \\
$rms$   &  0.07 &  0.06   &  --              & --               \\
\hline
$j_1$   &  1.37 $\pm$ 0.03 &  2.18 $\pm$ 0.03 & 1.91 $\pm$ 0.04 &  1.39 $\pm$ 0.04 \\
$j_2$   &  0.99 $\pm$ 0.01 &  0.93 $\pm$ 0.01 &  0.94 $\pm$ 0.01 & 0.99 $\pm$ 0.01 \\
$j_3$   &  -0.06 $\pm$ 0.01 & -0.02 $\pm$ 0.01 &  -0.03 $\pm$ 0.01 & -0.06 $\pm$ 0.01 \\
$rms$   &   0.03            &   0.04        &  0.04      &  0.04      \\
\hline
$h_1$   &   1.11 $\pm$ 0.03 &   2.71 $\pm$ 0.04 &  2.74 $\pm$ 0.04 &  1.76 $\pm$ 0.04 \\
$h_2$   &  1.01 $\pm$ 0.01 &   0.87 $\pm$ 0.01 &  0.88 $\pm$ 0.01 &  0.95 $\pm$ 0.01 \\
$h_3$   &  -0.01 $\pm$ 0.01 &   0.02 $\pm$ 0.01 &  0.02 $\pm$ 0.01 &  0.01 $\pm$ 0.01 \\
$rms$   &   0.03  &   0.04 &   0.04 &  0.05           \\
\hline
$k_1$   &   2.27 $\pm$ 0.04 &   4.25 $\pm$ 0.04 &  4.64 $\pm$ 0.03 &  2.60 $\pm$ 0.03 \\
$k_2$   &   0.98 $\pm$ 0.01 &  0.82 $\pm$ 0.01 &  0.78 $\pm$ 0.01 &   0.94 $\pm$ 0.01 \\
$k_3$   &   0.01 $\pm$ 0.01 &  0.01 $\pm$ 0.01 &  0.00 $\pm$ 0.01 &  0.01 $\pm$ 0.01 \\
$rms$   &  0.04 &  0.04 &   0.04    &  0.05           \\
\hline
\label{tab:frames2}
\end{tabular}
\end{table*}

\subsubsection{Mid infrared data}
We cross$-$correlated our photometric data with mid IR data (on the regions where these data are available) with the purpose of detecting {\rm PMS} stars (see also Sect~\ref{sec:photometricdiag}).

We used then, the {\rm MSX} data \citep{ega03} and {\rm SST} data. {\rm MSX} mapped the galactic plane and other regions missed or identified as particularly interest by the Infrared Astronomical Satellite ({\rm IRAS}) at wavelengths of 4.29, 4.35, 8.28, 12.13, 14.65 and 21.3 $\mu m$. On the other hand, the Infrared Array Camera ({\rm IRAC}) on-board the {\rm SST} was used to obtain images in four channels (3.6, 4.5, 5, 8 and 8.0 $\mu m$) and the {\rm GLIMPSE} catalogue was produced using several of these photometric data points.

We used then the {\it all matches} option in {\rm ALADIN} since this option gave us all the counterparts within the searching radius. Following these criteria, we found about 10 counterparts in {\rm MSX} and {\rm GLIMPSE} data on each cluster region (see Sect. \ref{sec:clustersregions} for more details).

\subsubsection{Final catalogues}
\label{sec:catalogues}
We used the {\rm STILTS}\footnote{http://www.star.bris.ac.uk/~mbt/stilts/} tool to manipulate tables
and to cross-correlate the optical and IR data. We obtained then four catalogues with astrometric/photometric information of about 76000 objects 
covering approximately a FOV of $10\farcm0 \times 10\farcm0$ around each studied cluster (see Fig.~\ref{fig:charts}). 
The corresponding photometric errors ($BVIJHK$ bands) in these catalogues are those provided by $DAOPHOT$ and $DAOMASTER$ codes. 
The full catalogues are available in electronic form at the CDS website.

\subsection{Spectroscopic data}
\label{spectroc}
We performed spectroscopic observations over eight stars in the region of {\rm DBS}\,77 embedded cluster distributed in three slit positions. We used the infrared spectrograph {\rm SofI}\footnote{http://www.eso.org/sci/facilities/lasilla/instruments/sofi.html} on the NTT at La Silla Observatory ESO, Chile in long-slit mode, in 2011 during the night of April 17$^{th}$. Using the medium resolution grism in the 3$^{rd}$ order, we covered the whole $K_S$ band, 2.00 - 2.30 $\mu$m, with a resolution of $\Delta \lambda$ 4.66 $\AA \cdot pix^{-1}$. We used a 1 arcmin slit, in order to match the seeing, which gives a resolving power of $R$~$\simeq$~1320. For optimal subtraction of the atmospheric OH emission lines, we used 15 sec (ID stars: 5; 8 and 3; slit 1); 100 sec (ID stars: 17 and 27; slit 2) and 150 sec (ID stars: 103; 21; and 39; slit 3) while nodding along the slit in an ABBA pattern: the star was observed before (A) and after (B) a first nod along the slit, then at position B a second time before returning to position A for a last exposure. The S/N for each spectrum is given in Table \ref{tab:sptphot}. As a measure of the atmospheric absorption, we observed bright stars of spectral type G. We used the standard procedures to reduce the data. For more details of our reduction procedures see \citet{moo98} and \citet{che12, che13}.

\begin{figure}[!ht]
  \centering
\includegraphics[width=0.4\textwidth]{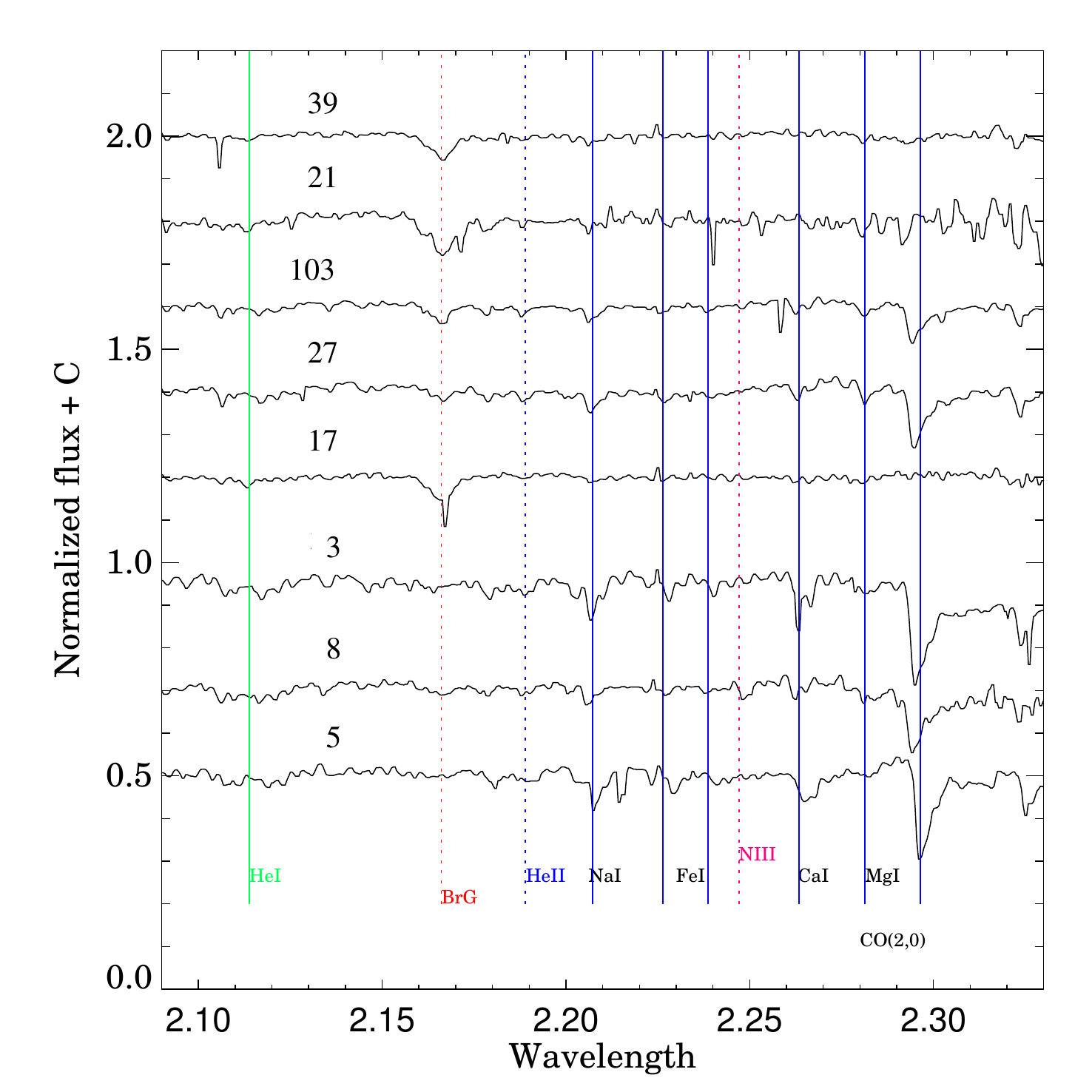} \\
 \caption{Stellar spectra in the {\rm DBS}\,77 region. Several chemical components of each star are marked with vertical lines. The spectral type of each star is indicated in Table \ref{tab:sptphot}}
  \label{DBS077SP}
\end{figure}    
\subsection{Radio data}
\label{radiodata}
With the aim to investigate the interstellar medium in which the star clusters are forming, we consulted the HI {\rm SGPS} data from the Australia Telescope Compact Array ({\rm ATCA}) and the {\rm PMN} survey carried out with the Parkes 64 m radio telescope and the {\rm NRAO} seven-beam receiver.

The {\rm SGPS} is a survey of 21$-$cm H\,I spectral line emission and continuum. The H\,I data were obtained using a channel separation of {\it $\Delta$v} $=$ 0.82 km s$^{-1}$ and the final $rms$ noise of a single profile is $\sim$~1.6~K on the brightness temperature ($T_b$) scale. The sensitivity of the radio continuum data is below 1 mJy beam$^{-1}$ and the angular resolution of the {\rm SGPS} is $1\farcm6$.

The 21$-$cm HI spectral line survey provides data covering 253$^{\circ}$ $\leq$ {\it l} $\leq$ 358$^{\circ}$ and $-$1$\fdg$5 $\leq$ {\it b} $\leq$ $+$1$\fdg$5, and a small region in the first Galactic quadrant between 5$^{\circ}$ $\leq$ {\it l} $\leq$ 20$^{\circ}$ and $-$1$\fdg$5 $\leq$ {\it b} $\leq$ $+$1$\fdg$5 ({\rm SGPS}\,I and {\rm SGPS}\,II, respectively, see \citet{mcc05} for details). The continuum survey only provides data of the 253$^{\circ}$ $\leq$ {\it l} $\leq$ 358$^{\circ}$ and $-$1$\fdg$0 $<$ {\it b} $<$ $+$1$\fdg$0 region (see \citet{hav06} for details). We processed the {\rm SGPS} continuum data using the {\it NOD2} package \citep{has74}. 

The {\rm PMN} is a southern sky survey at a frequency of  $\nu$~=~4.8 GHz. The angular resolution is 5$^{\arcmin}$ and the $rms$ sensitivity is $\sim$~8~mJy~beam~$^{-1}$. The survey provides several parameters (flux density, coordinates, angular size, orientation, etc.) of the H\,I distribution for $-$87$^{\circ}$ $\leq$ $\delta$ $\leq$ 10$^{\circ}$.

\section{Analysis} 
\label{sec:analysis}
Employed methods for determining the fundamental parameters of the young stellar clusters, (sizes, extinctions, distances, ages, members and masses) have already been described in several previous papers \citep[see e.g.][]{che12, che13, bau14}. 
Here, we provide a brief summary and highlight what we have done differently in this study.

\subsection{Centers and sizes}
The selected clusters are very faint and apparently have few members that makes difficult to
apply the radial density profile method. We then considered as the center coordinates, those given in the {\rm DBS}\, catalog as an initial point and the visual inspection of the infrared images together with the resulting photometric diagrams using different centers and radii. We finally adopted those values that contain the brightest stars and revealed the presence of the probable PMS population in these diagrams (see Table \ref{tab:param}).

\subsection{Kinematic study}
Proper motions and radial velocities are very useful tools to assign cluster memberships.
Unfortunately, the {\rm VVV} database currently contains observations covering four years and thus it is possible to determine the proper motions only for nearby Sun objects. On the other hand, the low resolution of the spectra, combined with the small wavelength coverage, does not allow a very accurate measurement of the radial velocities. 
Our data were then not reliable to provide membership probability from the kinematic behavior of the objects.

\begin{figure}[!ht]
  \centering
\includegraphics[width=0.3\textwidth]{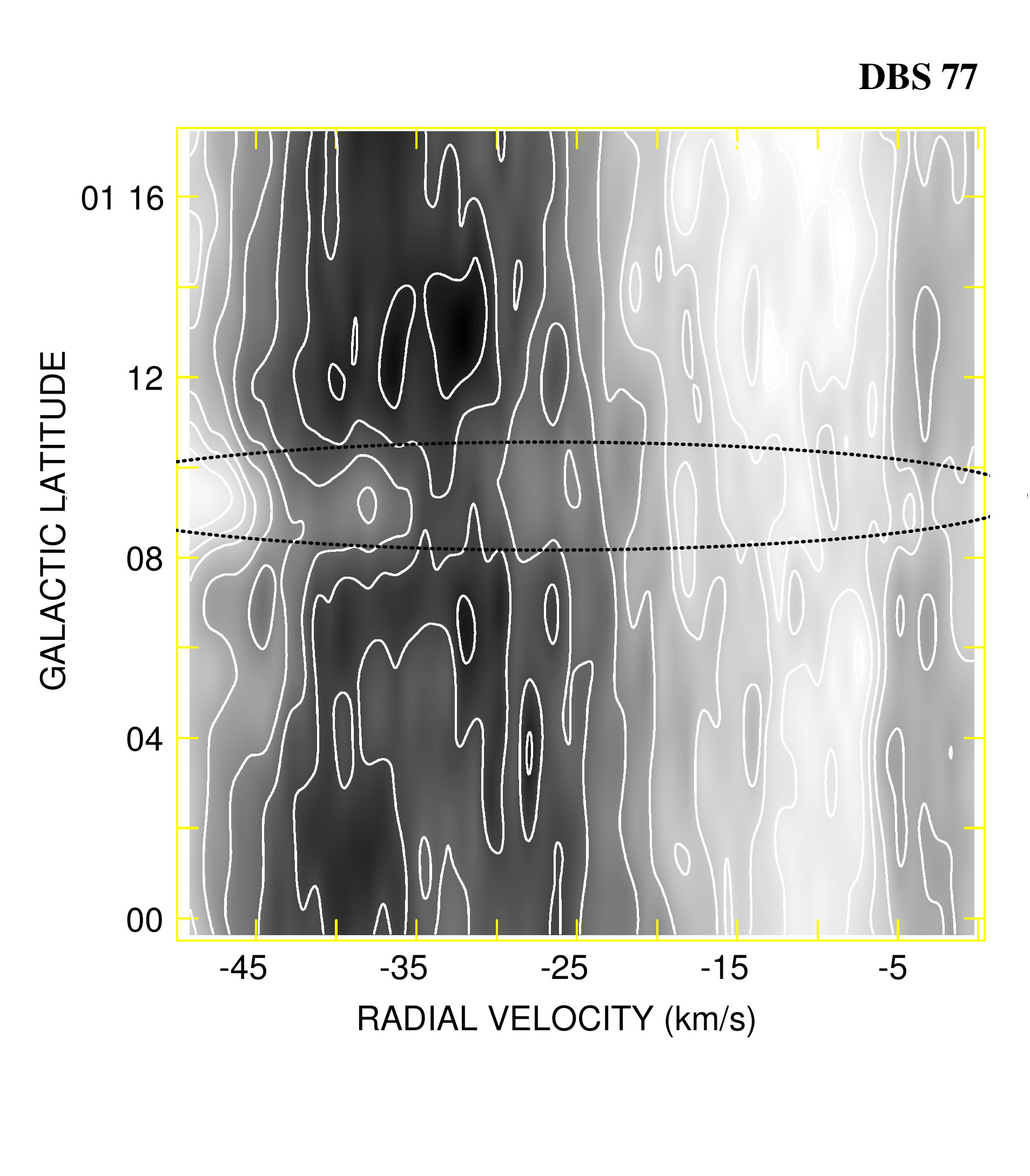} 
\caption{$T_b$ H\,I distribution at {\rm DBS}\,77 region ($l \sim 301^{\circ}$). Black curved lines indicate the place of the H\,II region ({\it RV} $= -$ 46 km s$^{-1}$) and the H\,I absorption in the line of sight. Lowest and highest contours are 12.4 K and 124 K, respectively. Contour spacing is 6 K until 50 K and 12 K from there onwards. Lighter gray color corresponds to the lowest temperature value.}
  \label{DBS077RV}
\end{figure}

\begin{figure}[!ht]
  \centering
  \includegraphics[width=0.5\textwidth]{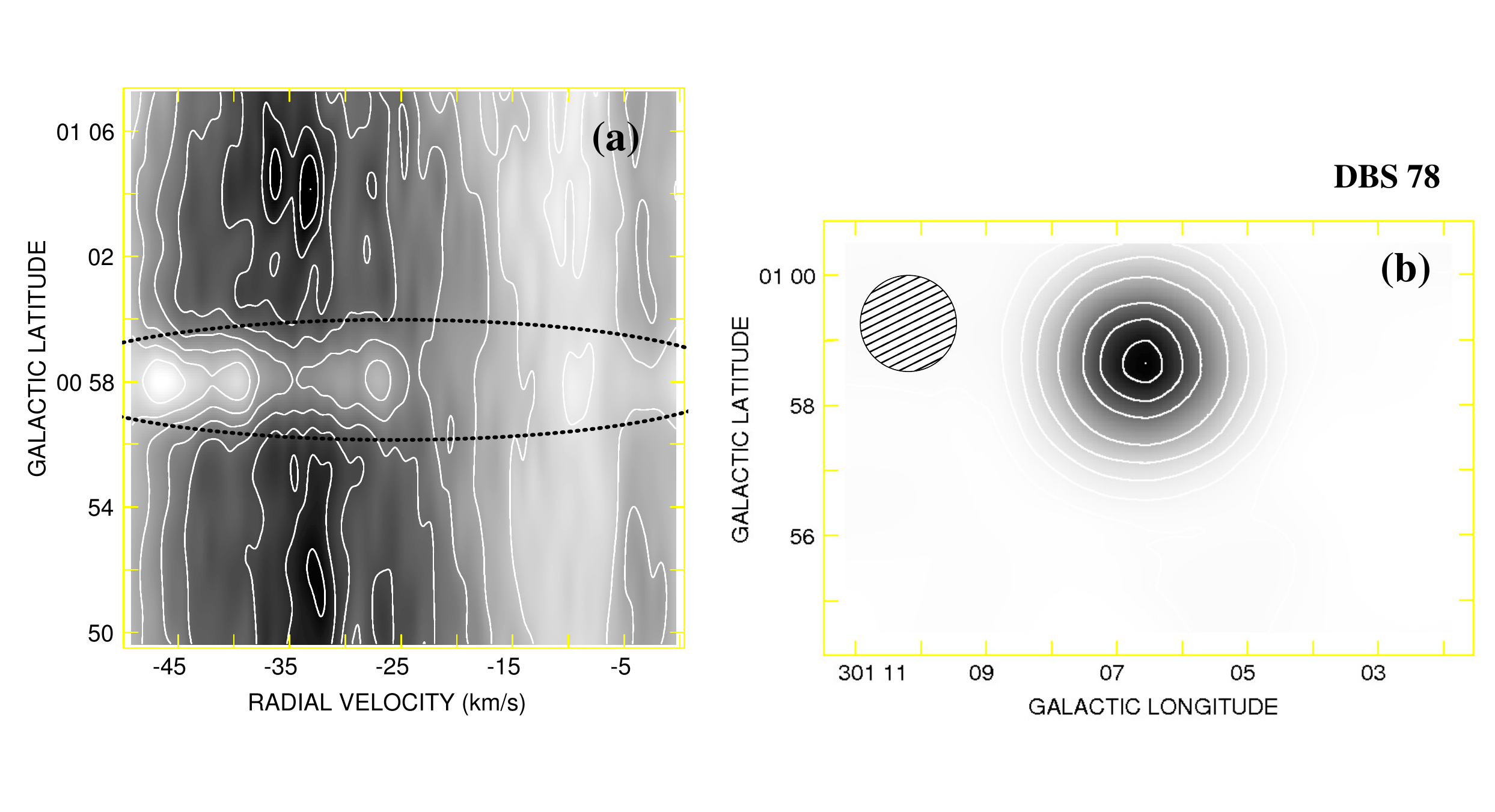} 
\caption{(a) Same as Fig.\ref{DBS077RV} but at {\rm DBS}\,78 region ($l \sim 301^{\circ}$). Lowest and highest contours are 15 K and 152 K, respectively. Contour spacing is 15 K. (b) Radio continuum emission at 1.42 GHz. Lowest and highest contours are 0.018 and 2.6 Jy beam$^{-1}$, respectively, without a constant step. Lighter gray color corresponds to the lowest flux value. Dashed pattern circle drawn in the upper left-hand corner indicates the angular resolution of the H\,I data.}
  \label{DBS078RV}
\end{figure}

\begin{figure}[!ht]
  \centering
  \includegraphics[width=0.5\textwidth]{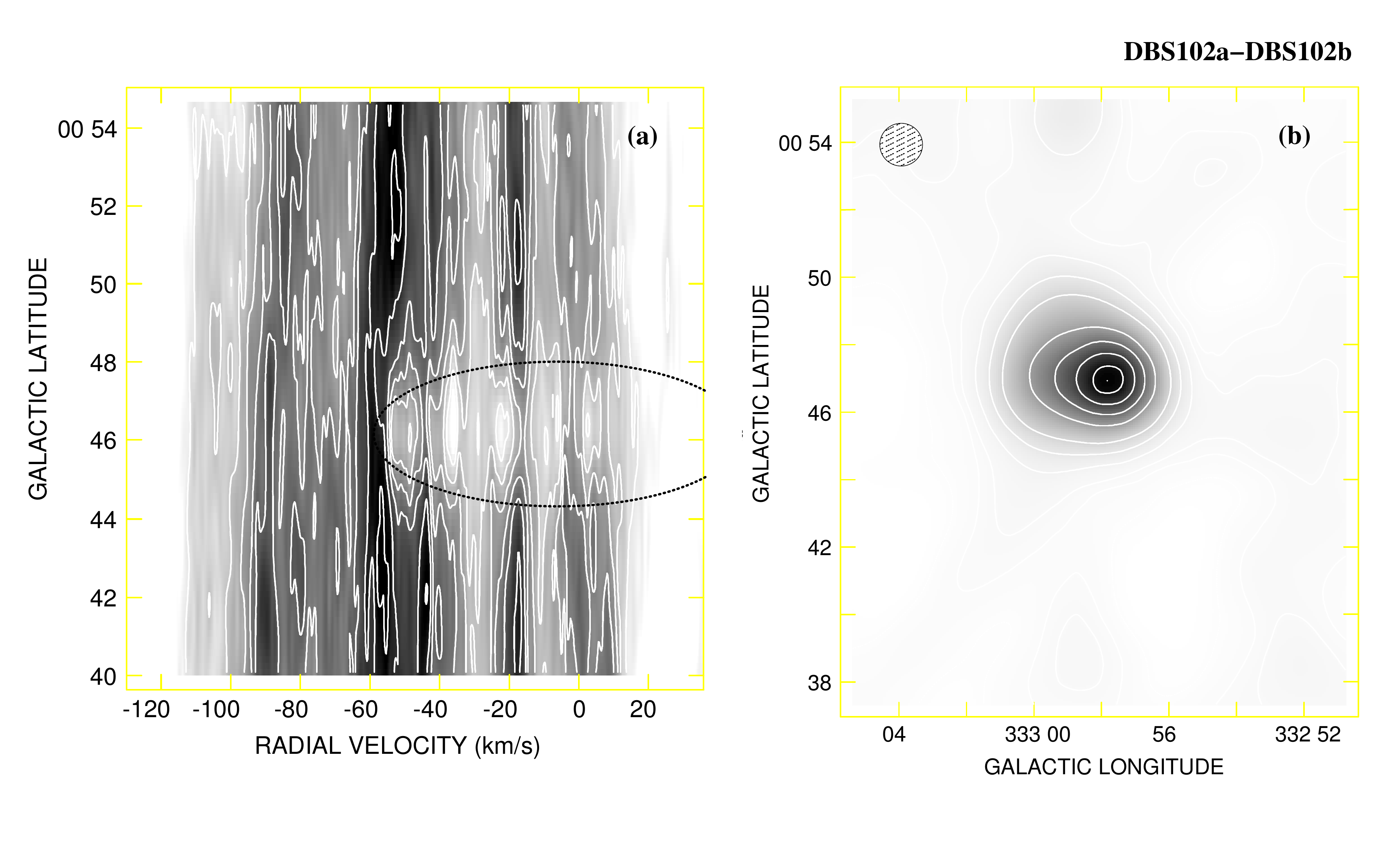} 
\caption{(a) Same as Fig.\ref{DBS078RV} but at  {\rm DBS}\,102 region ($l \sim 333^{\circ}$). Here the H\,II region has {\it RV} $= - $52 km s$^{-1}$. Lowest and highest contours are 30 K and 114 K, respectively. Contour spacing is 12 K. (b) Lowest and highest contours in radio continuum emission are 0.1 and 2.7 Jy beam$^{-1}$, respectively, without a constant step.}
  \label{DBS102RV}
\end{figure}

\begin{figure}[!ht]
  \centering
  \includegraphics[width=0.5\textwidth]{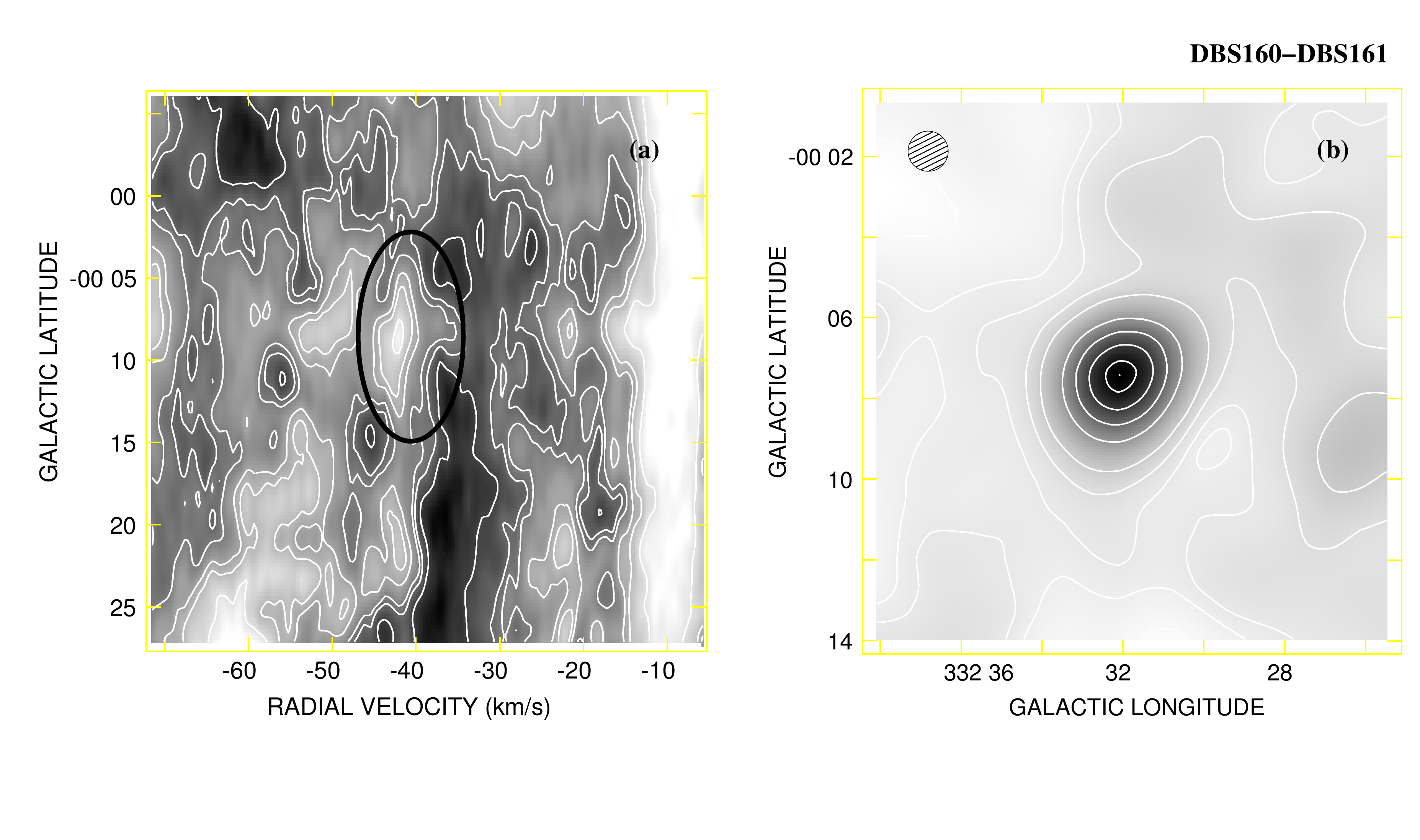} 
\caption{(a) Mean $T_b$ H\,I distribution at {\rm DBS}\,160/161 region between the longitude range from $332\fdg$52 to $332\fdg$55. Contour levels are from 65 K to 105 K in steps of 10 K. The ellipse indicates the interstellar bubble (Fig.\ref{fig:bubble}). H\,II region ({\it RV} $\simeq - $47 km s$^{-1}$) is not visible \citep{bro96}. (b) Same as Fig. \ref{DBS102RV} with the lowest and highest contours indicated 0.04 and 1 Jy beam$^{-1}$, respectively, without a constante step.}
 \label{DBS160RV}
\end{figure}

Under this situation, we made use of kinematic information of H\,II regions, molecular and dark clouds placed projected near ($\sim$ 10$^{\arcmin}$) each cluster. We transformed their radial velocities (RV), referred to the Local Standard of Rest ({\rm LSR}) to kinematic distances. In this procedure we used the Adjusted Linear Model (AL) and the Law Potential Model (LP) given by \citet{fic89} or the Galactic rotation model of \citet{bra93} (see Sect. \ref{sec:clustersregions}). The error in the determination of these kinematic distances arises due to an uncertainty of 10 km s$^{-1}$ due to non circular motion \citep{bur88}.
The H\,I maps, $RV$ vs $b$ (Figs. \ref{DBS077RV} to \ref{DBS160RV}), for all the studied regions allowed us to resolve the distance ambiguity problem present in the fourth Galactic quadrant.
The obtained kinematic distances yield us to have an starting point to analyze our photometric diagrams (see Sect. \ref{sec:photometricdiag}). 

\subsection{Spectrophotometric analysis}
\label{sec:spectroanalysis}
Spectrophotometric color excesses and distances were calculated for the stars for which we know their spectral classification from our spectra or from the bibliography. Our spectroscopic data involve some of the brightest stars of {\rm DBS}\, 77 region and we used them to determine their spectral classification, verifying their cluster membership situation. 
Spectral classification was performed using atlases of $K-$band spectra that feature spectral types stemming from optical studies \citep{ray09, lie09,  mey98, wall97} (see Fig. \ref{DBS077SP}).

We used the optical ($BVI$) and infrared ($JHK$) data and the calibrations given by \citet{lando82} and \citet{koo83}, respectively. We additionally evaluated the absorptions values on $V$ and $K$ bands as: $A_V = 3.1\,E_{B-V}$ and $A_K = 0.674 \,E_{J-K}$, valid for a normal reddening law (see Sect. \ref{sec:photometricdiag}). For each cluster, we selected the stellar members for which we are confident of their spectral types. We estimated individual stellar distances through the mean value of the computed $V_0 - M_V$ and $K_0 - M_K$ values. 


\subsection{Photometric diagrams}
\label{sec:photometricdiag}
\begin{figure*}[!ht]
\centering
\begin{tabular}{cc}
\multicolumn{2}{c}{\includegraphics[height=6.2cm]{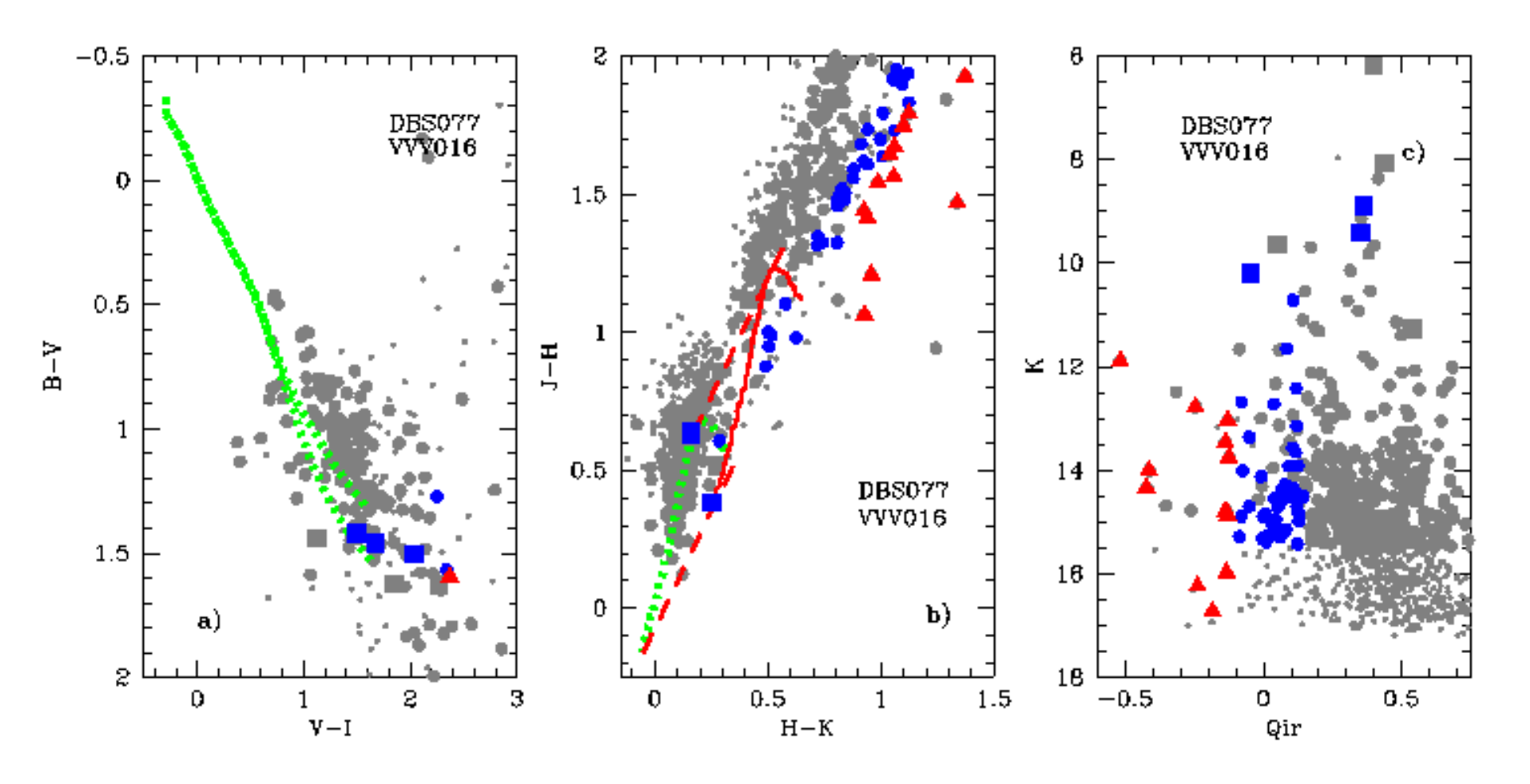}} \\
\includegraphics[height=6cm]{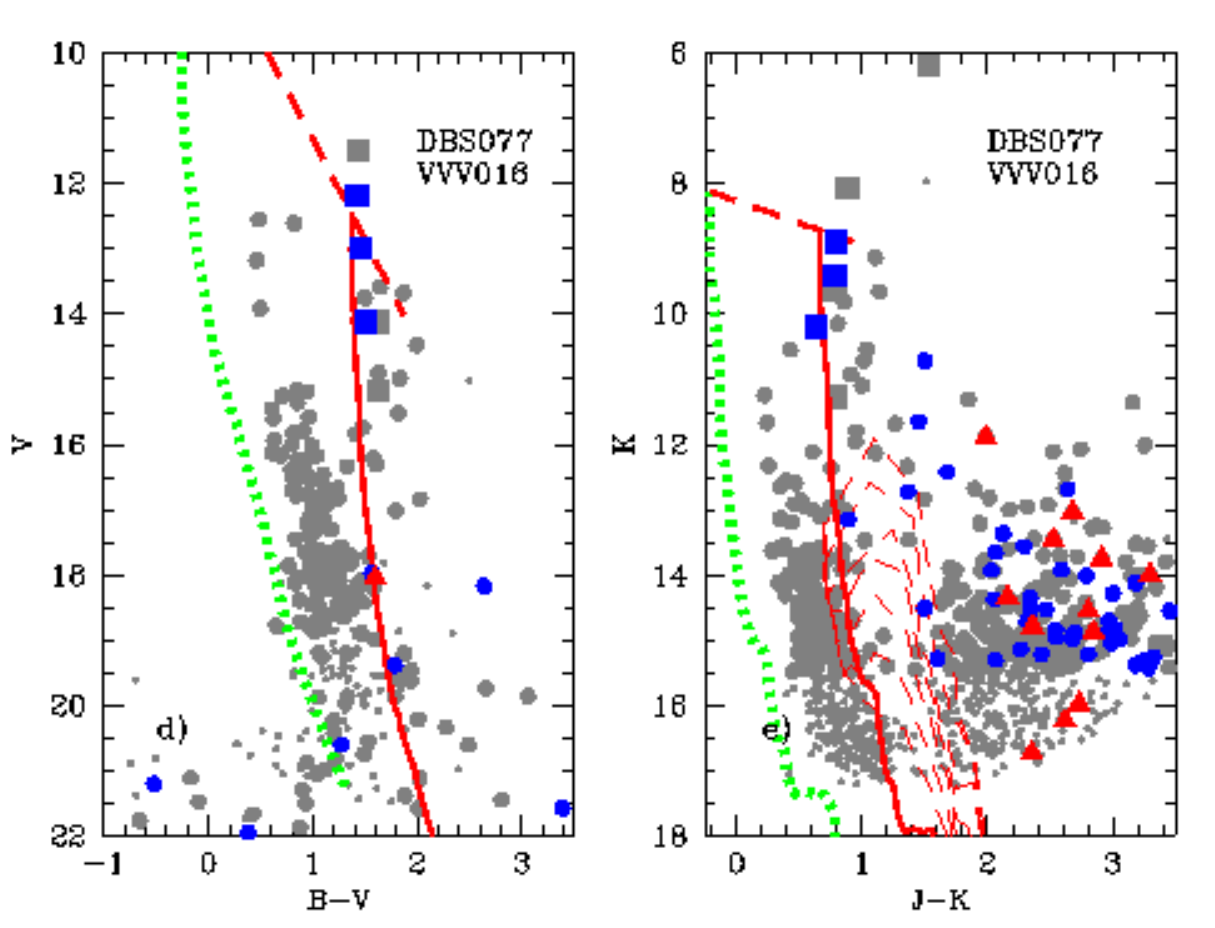} & \includegraphics[height=6cm]{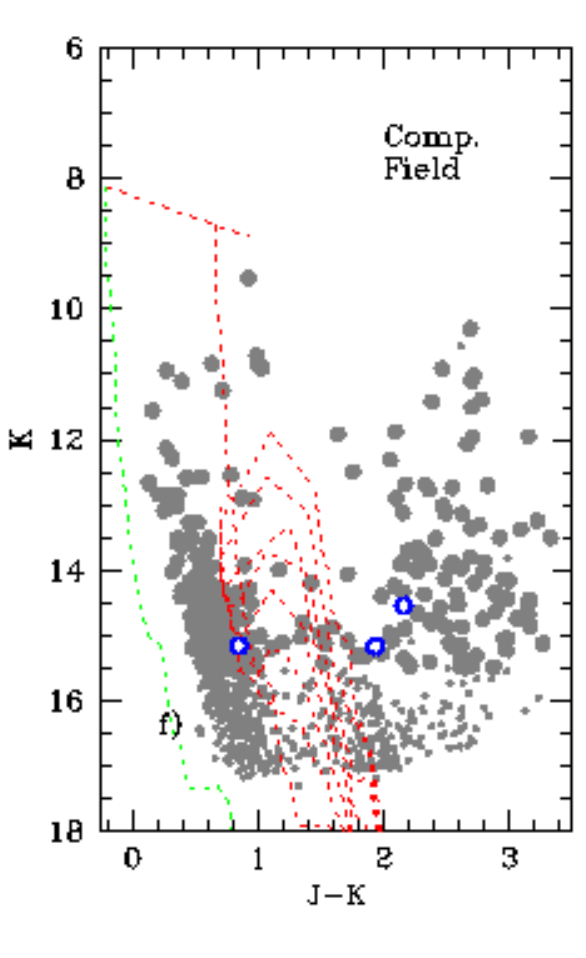} \\
\end{tabular}
\caption{Photometric diagrams for the stars in {\rm DBS}~77 + VVV~16 region and their corresponding comparison field (panel f). Squares and circles represent stars with and without known spectral classification, respectively. Light grey symbols indicate field population whereas blue and red represent different cluster populations selected by our photometric method (see text in Sect. \ref{sec:photometricdiag}). The solid (red) and dotted (green) curves are the ZAMS or MS (see text) shifted the adopted distance moduli with and without absorption/reddening, respectively. Dashed (red) lines indicate the normal reddening path ($R_V = 3.1$). Dashed curves are \citet{sie00} isochrones for $z = 0.02$.}

\label{fig:phot_dbs077}
\end{figure*}

\begin{figure*}[!ht]
\centering
\begin{tabular}{cc}
\includegraphics[height=6cm]{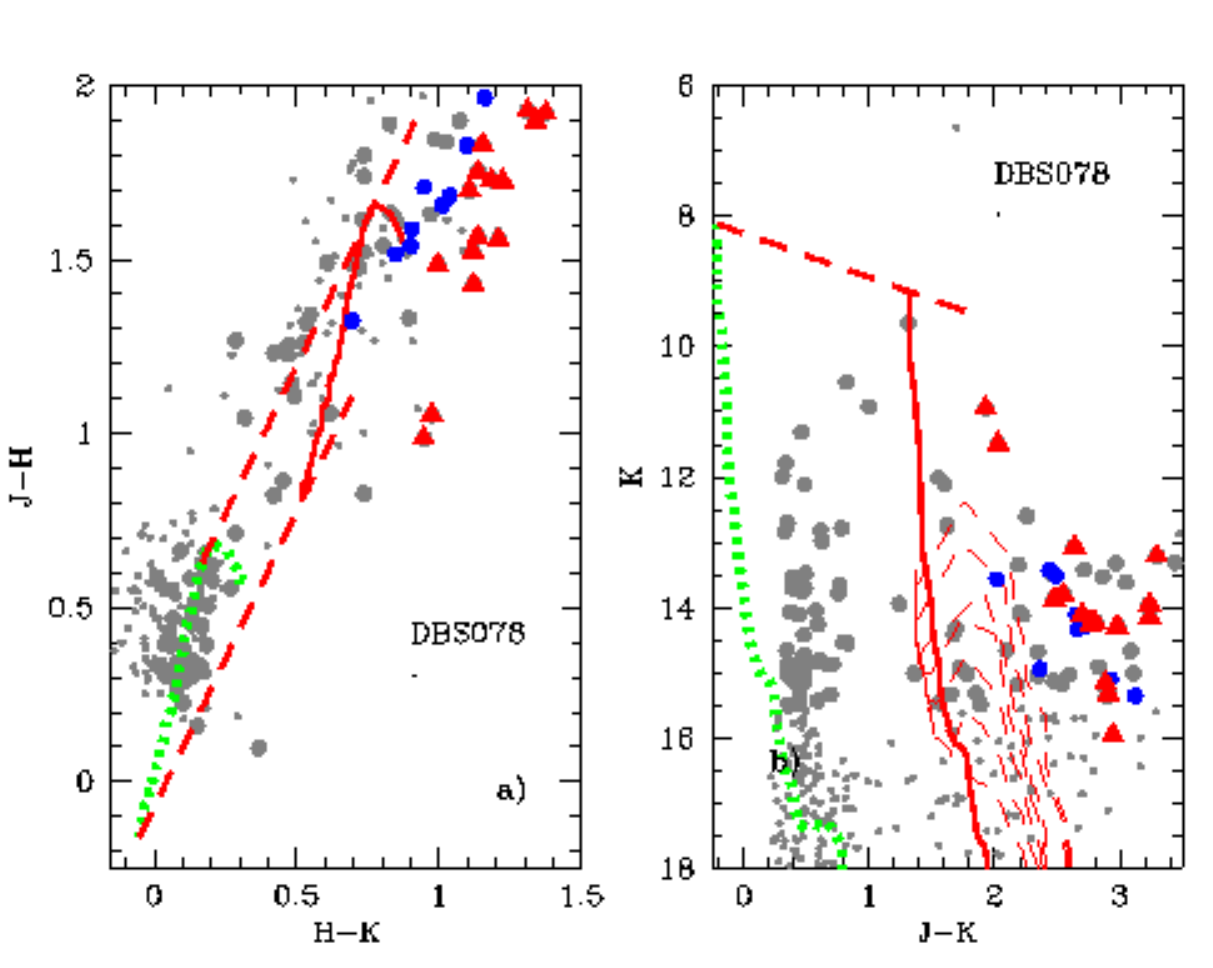} & \includegraphics[height=5.7cm]{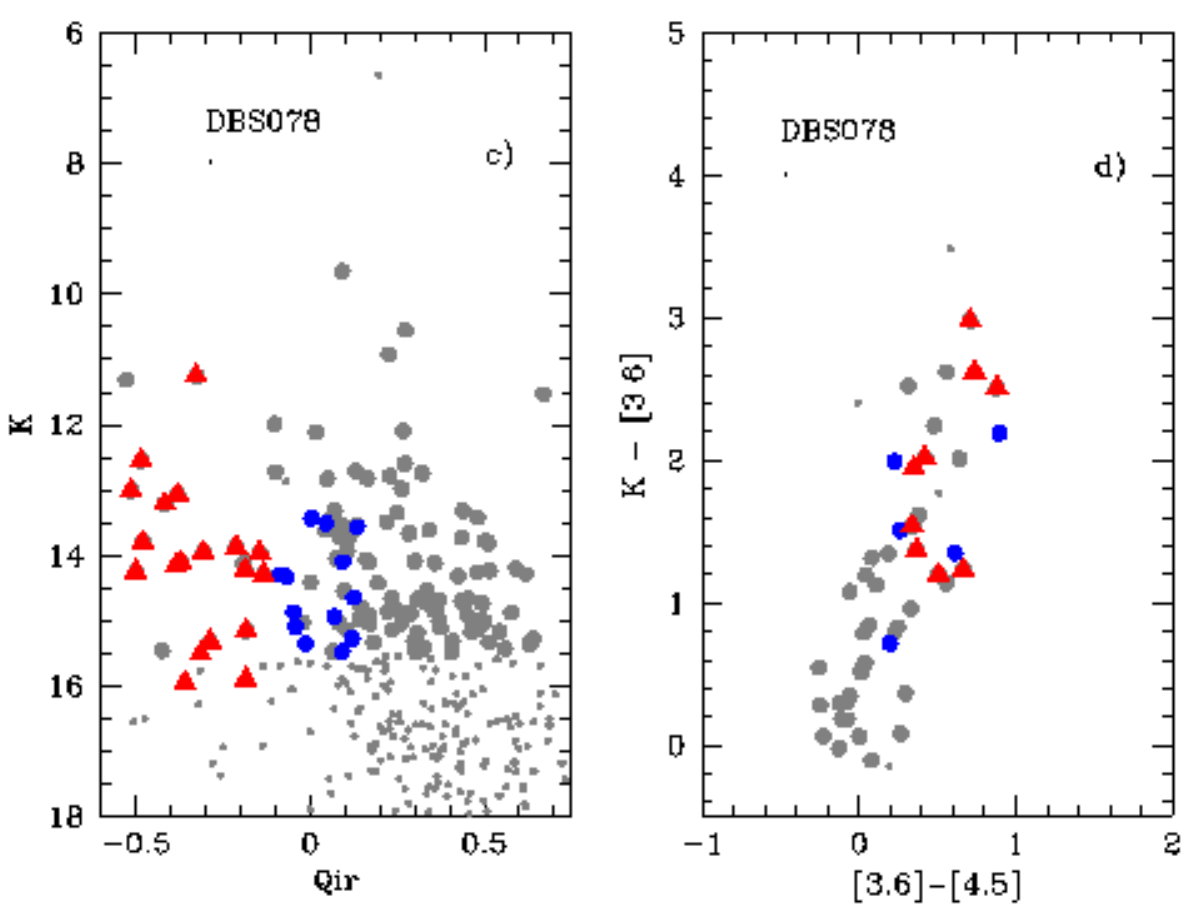} \\
\end{tabular}
\caption{Photometric diagrams for the stars in {\rm DBS}~78 region. Symbols as in Fig~\ref{fig:phot_dbs077}.}
\label{fig:phot_dbs078}
\end{figure*}

\begin{figure*}[!ht]
\centering
\begin{tabular}{cc}
\includegraphics[height=6.2cm]{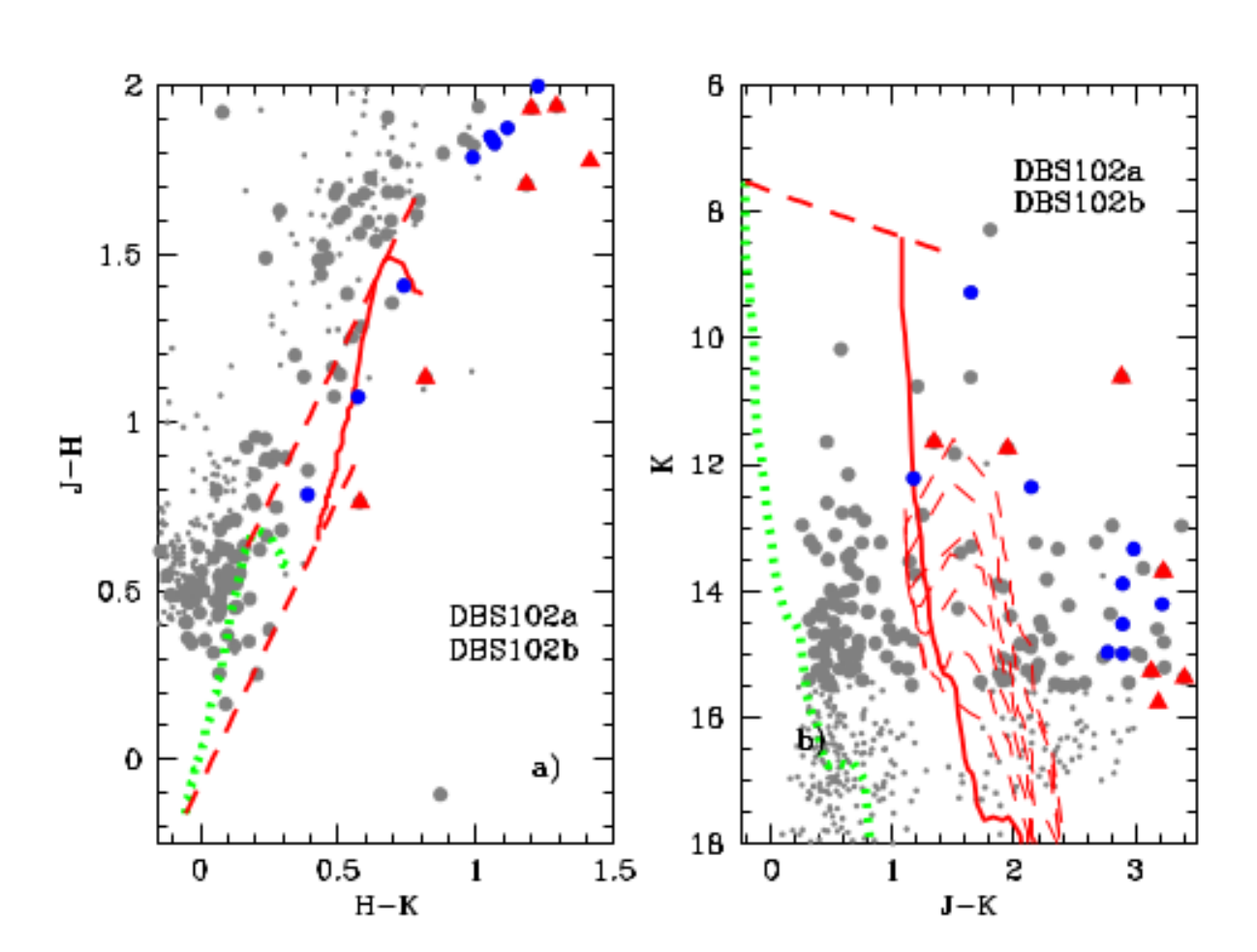} & \includegraphics[height=6cm]{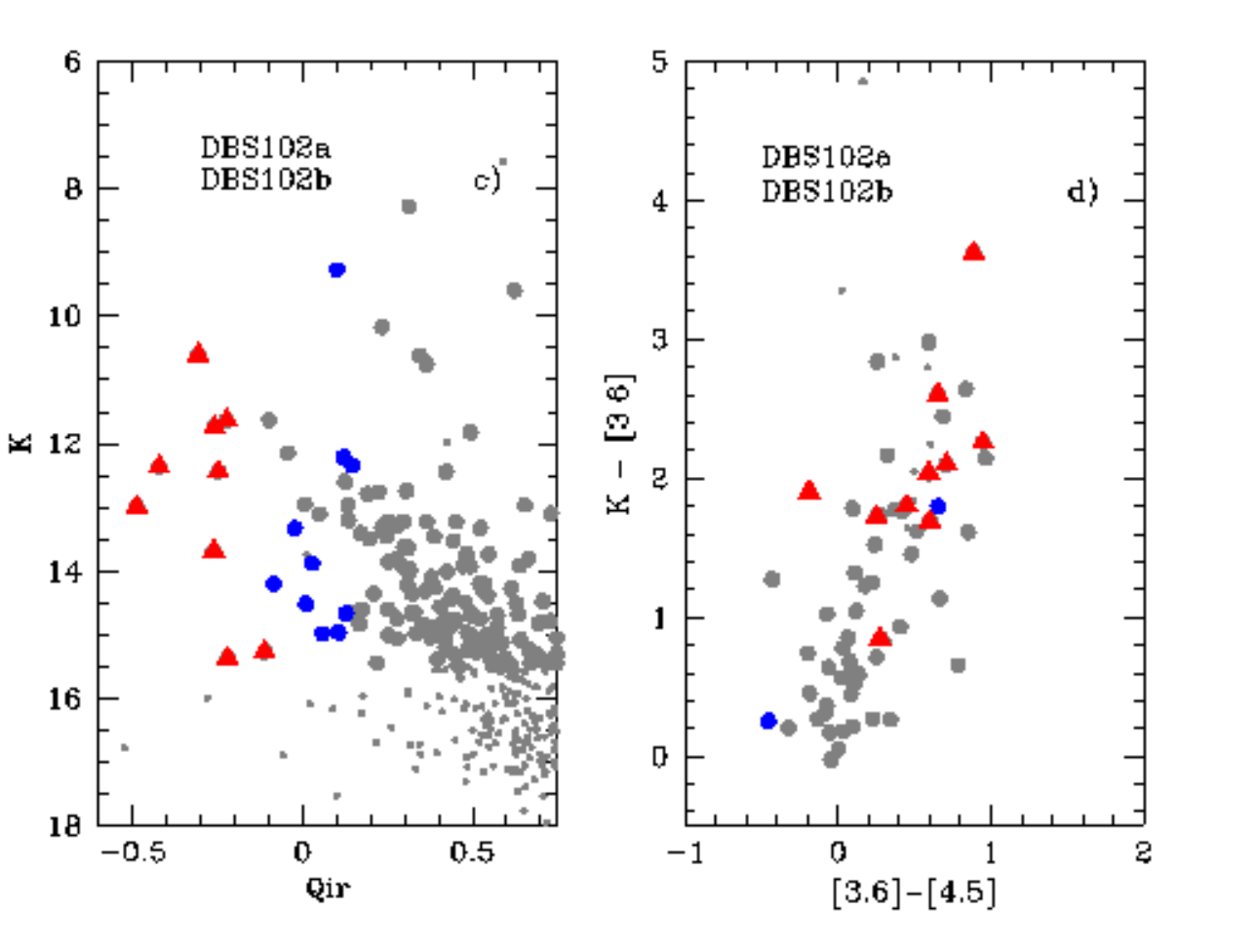} \\
\end{tabular}
\caption{Photometric diagrams for the stars in {\rm DBS}~102 region. Symbols as in Fig~\ref{fig:phot_dbs077}.}
\label{fig:phot_dbs102}
\end{figure*}

\begin{figure*}[!ht]
\centering
\begin{tabular}{cc}
\includegraphics[height=6.05cm]{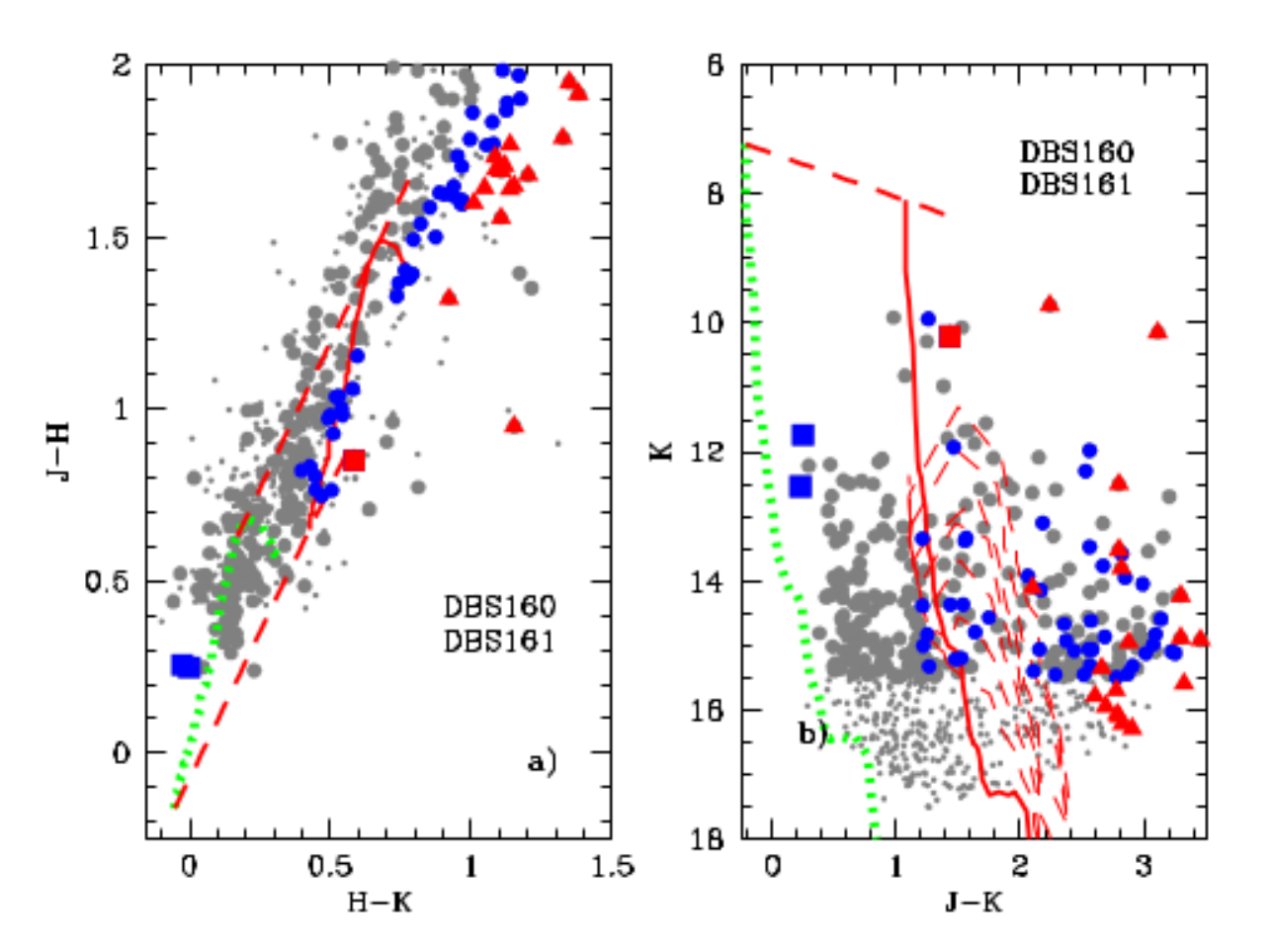} & \includegraphics[height=6cm]{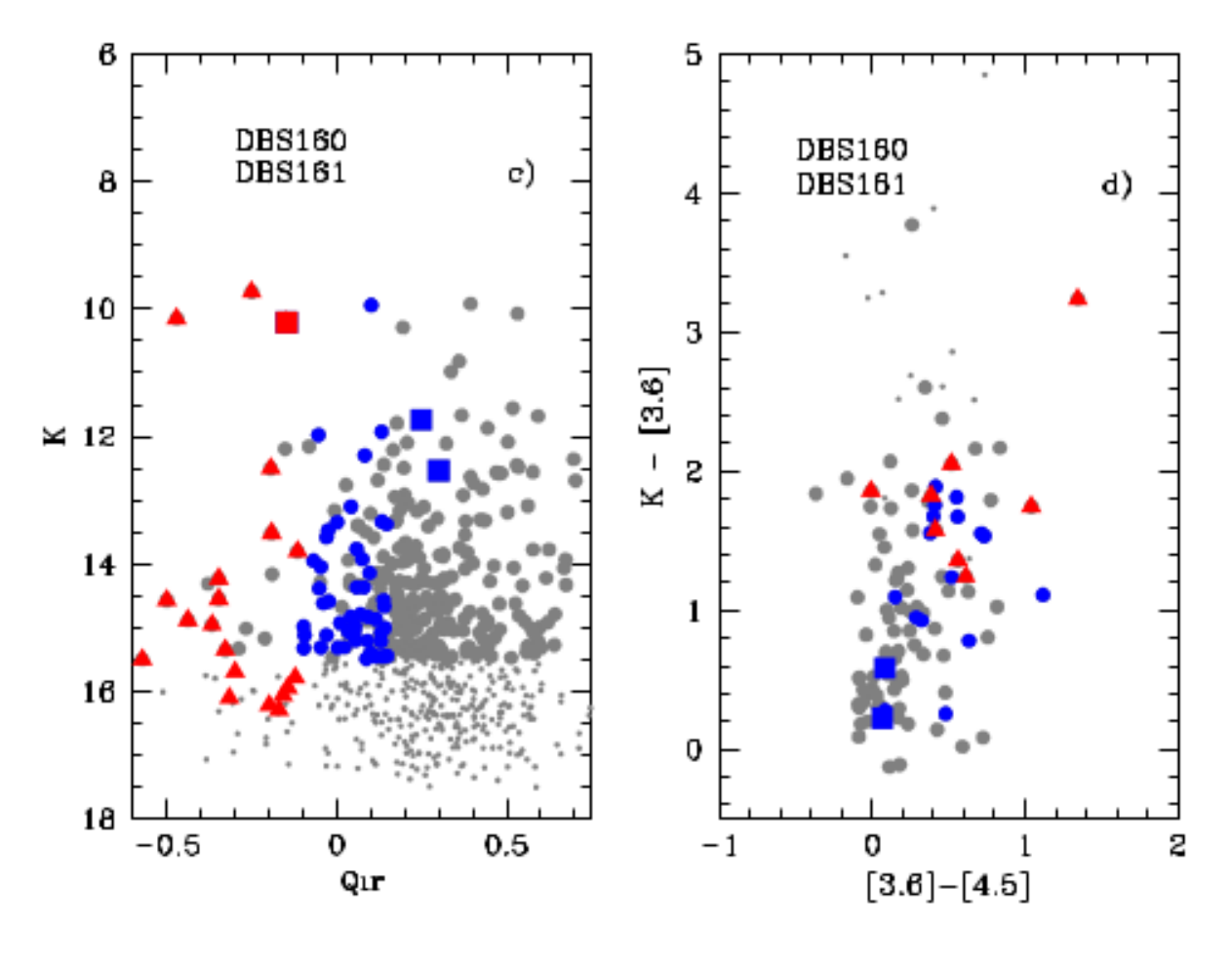} \\
\end{tabular}
\caption{Photometric diagrams for the stars in {\rm DBS}~160$-$161 region. Symbols as in Fig~\ref{fig:phot_dbs077}.}
\label{fig:phot_dbs160}
\end{figure*}

We checked the loci of the stars over all the possible color$-$magnitud diagrams and two color diagrams (\rm {CMD}s and \rm {TCD}s, respectively) obtained from our  catalogue and located inside the adopted clusters regions (see adopted centers and radii for each cluster in Table \ref{tab:param}). We present the corresponding diagrams along Figs. \ref{fig:phot_dbs077} to \ref{fig:phot_dbs160}. In particular, optical data are available for all the selected regions (except {\rm DBS}\,160$-$161 region). These data were used to check a consistent fit of the MS over all the diagrams. To save space, we only present the optical diagrams for the case of {\rm DBS 77} $+$ {\rm VVV} 16 region. In general, the photometric diagrams revealed the presence of two stellar populations:
a) the field population, usually late type MS stars, represented by the group placed at left in all de CMDs (grey symbols along the diagrams), and
b) the clusters populations suffering high and differential reddening and mixed with possible field red giants.
To separate both populations, we used (when available) the spectrophotometric analysis previously described, obtaining the corresponding color excesses and spectrophotometric distances.
For stars without spectral classification, we used all the photometric diagrams and the reddening$-$free photometric parameter $Q_{NIR} = (J - H) - 1.7 (H - K)$ \citep{neg07} in order to avoid the intrinsic degeneracy between reddening and spectral type. In particular, the $(J-H)$ vs. $(H-K)$ and/or the $K$ vs. $Q_{NIR}$ diagrams could be used to distinguish different stellar populations.

In our procedure, we adopted as cluster members those stars placed approximately redder than an adopted consistent MS (red curve) along all the photometric diagrams. We adopted the MS calibrations given by \citet{lando82}, \citet{cou78} and \citet{koo83}. Their locations were computed using the adopted distances and color excesses presented in Table \ref{tab:param} and a normal reddening behavior (see Sect. \ref{sec:spectroanalysis}) as was suggested by \rm{TCD}s diagrams. This fact allowed us to use the absorption ratios ($r_X = A_X/A_V$) given by \citet{joh68} van der Hulst curve 15 and \citet{cam02}. Then, following \citet{bor12} and \citet{mes12} we 
separated the selected stars among early normal MS stars ($-0.1 < Q_{NIR} < 0.15$; blue symbols) and, stars with emission features or revealing IR excess ($Q_{NIR} < -0.1$; red symbols), including for example PMS stars and YSOs. 
We noticed that some spectroscopically confirmed OB stars in DBS~77 and DBS~160 regions (see Figs. \ref{fig:phot_dbs077} and \ref{fig:phot_dbs160}) have high $Q_{NIR}$ values, but it could be a binarity effect. To  confirm the reliability of our procedure, we conducted a similar analysis over a 
comparison field region for each cluster. In these cases, as was expected, we found almost no objects that could be considered cluster members. It must be noticed, 
however, that we do not provide an unambiguous membership diagnostic, but we defined a uniform approach to select the likely member stars used to calculate the clusters 
parameters.

To estimate the absolute magnitudes and spectral types for the adopted \rm {MS} cluster stars without spectroscopic observations, we dereddened their location over the \rm {CMD}s following a normal path until their intersection with the \rm {MS}. We present in Table~\ref{tab:sptphot} our adopted values and computations for studied stars with spectroscopic data and those ones more relevant ($K < $13\, or\, 14 depending on the cluster).


\subsection{Individual stellar masses and {\rm IMF} computation}
\label{stellarmass}
We obtained the stellar mass for the adopted cluster members using the computed absolute magnitude values (see Sec. \ref{sec:photometricdiag}). We employed the ZAMS obtained from \citet{bre12, chen14, tang14, chen15} and we interpolated the stellar mass using the algorithm given by \citet{ste90}. This procedure is valid only for MS stars, and for adopted PMS stars is only provided an approximate upper estimation of the real mass values.

The initial mass function ({\rm IMF}) for massive stars can be modeled as a power law. This is:
\begin{equation}
\log(N/\Delta(\log{m})) = \Gamma \log{m}
\end{equation}
\noindent where {\it N} is the number of stars per logarithmic mass bin {\it $\Delta(\log{m})$}. Since the computed spectral type depends upon the used photometric bands, we obtained several possible mass values for each star. Therefore, to build the IMF, we adopted the mean mass value obtained from \citet{mar05} for O$-$spectral types and from \citet{pec13} for B and later spectral types.

\begin{figure}[!ht]
  \centering
  \includegraphics[width=0.4\textwidth]{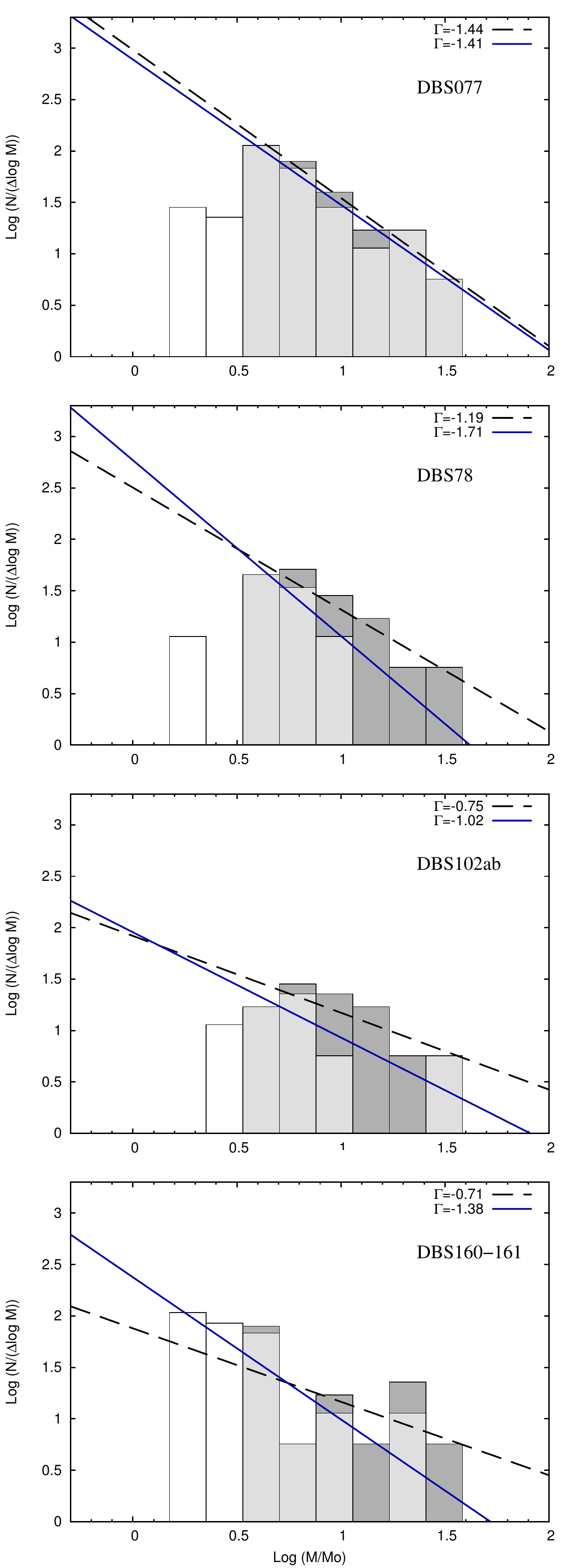} 
  \caption{Distribution of stellar masses for the studied embedded clusters. Grey bars (light and dark) indicate more massive bins employed for compute the IMF slopes ($\Gamma$). Light grey bars and blue solid lines correspond to only MS stars, whereas dark grey bars and black dashed lines correspond to MS$+$PMS stars.}
\label{Group300}
\end{figure}

We built two IMFs for each studied stellar group taking into account only the adopted MS stars and adding the adopted PMS stars. It must be noticed 
that the uncertainly in their mass values was minimized by the used width of the bins in the IMF estimation. Our results 
are presented in Fig. \ref{Group300}, together with the least squares fit for most massive stars ($>$ 3 M$_{\odot}$), avoiding the very probable incomplete less massive bins. We noticed that, except in the case of {\rm DBS}\,77 region, the presence of PMS has an important effect on the obtained slope values (see Table~\ref{tab:param} and Fig. \ref{Group300})

\subsection{Estimating cluster ages}
\label{ages}
Since all the clusters presented strong differential reddening it was very difficult to estimate their ages using the classical method of fitting the observed CMD with isochrone sets.
We therefore took into account the early spectral type of the stars adopted as cluster members and the \citet{eks12} evolutionary models to calculate the MS duration from the mass of the star. In case where spectroscopic data were not available, we used the estimated spectral classification from the photometric data (see Sect.\ref{sec:photometricdiag} and Table \ref{tab:sptphot}). It must be noticed that the estimated age values are upper limits since the reveled high amount of PMS candidates suggest an active star formation process in all the studied clusters.

\subsection{Studying the radio maps}
 \label{sec:radiomaps}
The selected radio data (see Sect. \ref{radiodata}) provide H\,I spectral line information over all the selected regions, but unfortunately, the continuum survey only provides data until $b = 1^{\circ}$, for this reason we could not analyze the H\,I distribution at 1.4 GHz for {\rm DBS}\,77 region.
We derived the angular diameters and flux densities ($S_{\nu}$) of the radio continuum point sources found in the direction of {\rm DBS}\,160$-$161 and {\rm DBS}\,78 using a bidimenssional gaussian fit to the SGPS continuum data. The error in the fit was about 2 \% and we determined a $rms$ flux noise level of 0.013 Jy beam$^{-1}$ from a source free region located in the neighborhood. 
The obtained fitted parameters are presented in Table \ref{tab:HIIparam}.
 
In the direction of {\rm DBS} 102 we found an extended source (for the angular resolution of the SGPS survey), with an angular size of 7$\farcm$3 $\times$ 6$\farcm$9.
We calculated the integrated flux density of the source using the method of polygon integrations and we subtracted the background with a lineal baseline fit. The quoted error reflected the uncertainties in determining the background levels, which added up a total of about 5\%.

The emission mechanism of the detected radio sources could be identified using the spectral index $\alpha$ ($S \sim \nu^{-\alpha}$). This index was  obtained from the measured flux density at two different frequencies, $\nu_1$ and $\nu_2$ ($\alpha = log(S_1 / S_2)/log({\nu_2}/{\nu_1})$). 
We computed the corresponding spectral indices for the radio sources found at the direction of the clusters {\rm DBS}\,78, {\rm DBS}\,102, and {\rm DBS}\,160-161 using the {\rm SGPS} data (1.4 GHz) and the {\rm PMN} data (4.85 GHz). The $\alpha \simeq$ 0.1 values found with both frequencies revealed a spectra whose behavior was expected for an optically thin plasma. The obtained values are presented in Table \ref{tab:HIIparam}. 

The bidimensional gaussian fit and the method of polygon integration (for point or extended sources, respectively) also provided the peak brightness temperature ($T_C$) for each source at 1.4 GHz. This parameter ($T_C$), for these regions, was also computed by \citet{brow14} obtaining different values, however they employed the "on-off method" (see \citealt{kol03}). The $T_C$ values obtained by us and by \citet{brow14} are shown in Table \ref{tab:HIIparam}. 

The analysis of the observations in the radio continuum spectrum of the optically thin nebula allowed us to estimate the excitation parameter ${\it u}\,[pc\,cm^{-2}]= R_s N^{2/3}_e$. This value represents the Lyman photon number that might be absorbed by the H\,II region to be ionized.
The linear Str\"{o}mgren radio, $R_s$, was obtained as the geometric mean of the angular size of the source and $N_e$ is the electronic density. 
This latter value was computed for each region using a geometric model and employing the expression given by \citet{mez67}:
\begin{eqnarray}
\label{Ne}
\nonumber N_e[cm^{-3}] = c_1 a^{1/2} 635.1 (T_e [10^4k])^{0.175} (\nu[GHz])^{0.05} (S_{\nu}[Jy])^{1/2}\\
(1/d[kpc])^{1/2}(\theta_G[arc min])^{-1.5}
\end{eqnarray}
\noindent where $c_1 = 0.775$ is the conversion factor employed for the spheric geometric model adopted for the nebula, $a = 1$ is the recombination coefficient at all levels, $T_e$ is the electronic temperature obtained from the model of \citet{qui06}, $S_{\nu}$ is the flux density measured in the H\,II region at $\nu =$ 1.42 GHz and $\theta_G$ is the observed apparent half-power beam width (\rm HPBW), obtained as:
\begin{equation}
\label{thetaobs}
\theta_{G} =  (\theta_{sph1} \theta_{sph2})^{1/2}/1.471
\end{equation}   
\noindent being $\theta_{sph1}$ and $\theta_{sph2}$ the geometric mean of the observed size of the source in two perpendicular coordinates.

Moreover, using models of stellar atmospheres it is possible to determine the total number ($N_{Ly}$) of ionizing photons in the Lyman continuum of the hydrogen emitted by the exciting star. We adopted the $N_{Ly}$ values given by \citet{smi02}. They calculated non-LTE, line-blanketed model atmospheres for several metallicities that cover the entire upper Hertzprung$-$Russell diagram for massive stars with stellar winds. 
We computed then $N_{Ly}$ for adopted cluster members stars earlier than B\,2.0, since they provide almost all these photons. 

Thus, the ionization parameter could be calculated as:
\begin{equation}
\label{photons}
U\,[pc\,cm^{-2}] = [3 N_{Ly} / 4 \pi a(2)]^{1/3}
\end{equation}
\noindent where $a(2) = 2.76 \times 10^{-13}[cm^3\, s^{-1}]$ represents the probability of recombination at all levels except the ground level.
We then compared the excitation parameter {\it u} with the parameter {\it U} calculated with Eq. \ref{photons} for all the exciting stars. Thus, we compared the observational results of density and radius of the nebula with the flow of ionizing photons emitted by the exciting stars. The comparison allowed us to determine whether the H\,II region deviates from the ideal conditions and what might be the causes. The obtained results of these comparisons are discussed in Sec. \ref{sec:HIIregions} and are presented in Table \ref{tab:HIIparam}.

An optically thin H\,II region also allowed us to estimate the mass of ionized hydrogen in the region following the expression given by \citet{mez67}.
\begin{eqnarray}
\label{MHII}
\nonumber M_{H\,II}[M_\odot] = c_2 a^{1/2} 0.3864 (T_e [10^4k])^{0.175} (\nu[GHz])^{0.05} (S_{\nu}[Jy])^{1/2}\\
(d[kpc])^{2.5}(\theta_G[arc min])^{1.5}
\end{eqnarray}
\noindent where $c_2 = 1.291$ is the conversion factor employed for the spheric geometric model adopted for the nebula and the other observable quantities are the same as those used in Eq. \ref{Ne}. 

The 21$-$cm line emission maps of the studied regions (see Figs. \ref{DBS077RV} to \ref{DBS160RV}) show zones with low values of $T_b$. In the case of {\rm DBS}\,77, {\rm DBS}\,78, and {\rm DBS}\,102 regions, these minima could be interpreted as the result of the absorption of the H\,I distribution indicating that $T_c > T_b$. On the other hand, the {\rm DBS}\,160$-$161 region, presents a minimum in the H\,I emission distribution surrounded by enhanced H\,I emission, revealing the presence of a bubble. Interstellar bubbles are cavities with a minimum in the center of its H\,I distribution expanding at relatively low velocities $(\leq 10\,km s^{-1})$ \citep{cap03}.
\begin{table*}
\fontsize{10} {14pt}\selectfont
\caption{Main parameters of the H\,II regions linked with the studied embedded clusters}
\centering
\begin{tabular}{lccc}
\hline\hline
$ID$ & {\rm DBS}\,$78$ &  {\rm DBS}\,$102$ & {\rm DBS}\,$160-161$ \\
\hline
(l,b) [$^{\circ}$] & ($301.11$, $+ 0.97$) & ($332.97$, $+ 0.78$) & ($332.54$, $-0.13$) \\
  $\theta_{sph}$ $[']$ &  $1.8$ & $7.1$ & $1.9$ \\
 $T_c$ $[K]$ & $162^{(i)}$  &  $24^{(ii)}$ & $57^{(iii)}$ \\
 $T_e$ $[10^3 K]$ &  $9.7 \pm 0.3$ &  $9.5 \pm 0.2$ & $8.4 \pm 0.1$ \\
 $S_{1420}$ $[Jy]$ &  $3.5 \pm 0.1$ &  $7.1 \pm 0.3$ & $2.0 \pm 0.0$ \\
$S_{4850}$ $[Jy]$ & $3.9 \pm 0.1$ & $6.6 \pm 0.1$ & $1.8 \pm 0.1$ \\
$R_S$ $[pc]$ & $1.3 \pm 0.3$ & $3.7 \pm 1.0$ & $0.9 \pm 0.3$ \\
$N_e$ $[cm^{-3}]$ & $307 \pm 34$ & $66 \pm 9$ & $249 \pm 38$ \\
 {\it u} $[pc \cdot cm^{-2}]$ & $56 \pm 12$ & $60 \pm 17$ &  $37 \pm 11$ \\
$M$ $[M_{\odot}]$ & $61 \pm 34$ & $350 \pm 244$ & $21 \pm 17$ \\
$ RV_{(LSR)}$ [km s$^{(-1)}$] & $-$46$^{(1)}$ & $-$52.0$^{(2)}$ & $-$46.7$^{(3)}$ \\
 $\alpha$ & $+0.09 \pm 0.03$ & $+0.06 \pm 0.03$ & $+0.10 \pm 0.03$ \\      
\hline
\label{tab:HIIparam}
\end{tabular}
\vspace{-0.6cm}
\tablefoot{
\tablefoottext{i}{$119 K$ \citep{brow14}} 
\tablefoottext{ii}{$114 K$ \citep{brow14}} 
\tablefoottext{iii}{$10 K$ \citep{brow14}} 
\tablefoottext{1}{$-$ 40.8 km s$^{(-1)}$ \citet{bro96}}
\tablefoottext{2}{\citet{kuc97}}
\tablefoottext{3}{\citet{bro96}}
}
\end{table*}

\section{Describing clusters regions}
\label{sec:clustersregions}
\subsection{{\rm DBS}\,77, {\rm VVV}\,15, and {\rm VVV}\,16}
Since these three clusters are placed at the same studied region and their photometric diagrams revealed similar properties, they were studied as only one stellar group. We performed spectral classification of eight stars in this region (see Sect. \ref{spectroc}), obtaining that three of them could be considered cluster members and gave us an spectrophotometric distance of 4.6 kpc (see Table \ref{tab:param}). The 21$-$cm line emission map in this region (see Fig. \ref{DBS077RV}) shows that the largest H\,I absorption signature corresponds with a velocity value of $-$ 46 km s$^{-1}$. This value in this Galactic region, has no distance associated from the kinematic rotation models of the Galaxy (see Fig. \ref{Group300}), that means that this velocity is a "forbidden value" for this models and the perturbations of the rotation of the Galaxy could be causing it.
The radial velocity of the region, according to the \citet{bra93} model would be approximately $-$ 33 km s$^{-1}$ which corresponds to the tangential point linked to a distance of $\simeq$ 4.6 $\pm$ 0.5 kpc. This value is similar to the spectrophotometric one previously indicated and is consistent with other kinematic determinations \citep{fic89,bro96}. Therefore, we adopted 4.6 kpc as the most reliable distance for the studied stellar group in this region.

Our photometric analysis indicated the presence of an important amount of early MS stars (4 O type stars and 40 B type stars) and, 13 probable PMS stars.
In this region we found 6 {\rm MSX} sources candidates to YSOs, \citep{mot07} with 8 possible IR counterparts. 

We could not calculate the excitation parameter ${\it u}$ of this region because there are no map with H\,I emission distribution data in the continuum at 1.4 GHz, for this region.

\subsection{{\rm DBS}\,78}
\begin{figure}[!ht]
  \centering
  \includegraphics[width=0.4\textwidth]{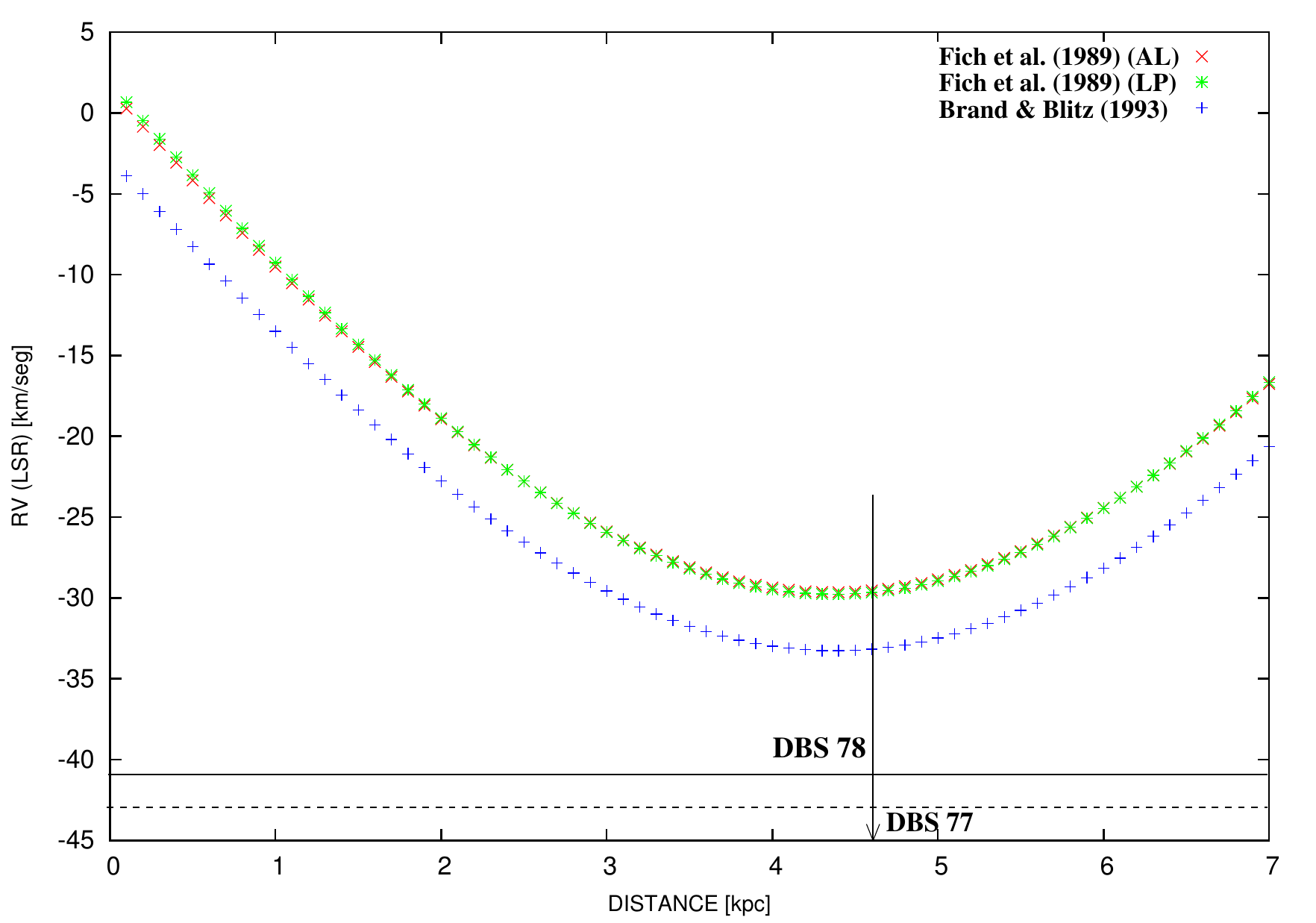} 
  \caption{Galactic rotation models applied at l = 301$^{\circ}$ to our Galaxy. Arrow indicates the adopted distance to the radial velocity of the radio sources according to the model.}
  \label{Grup300}
\end{figure}

There are no stars with spectral classification in this region, so we employed the 21$-$cm H\,I emission of the {\rm DBS}\,78 region to obtain its kinematic distance (see Fig. \ref{DBS078RV}). We found that the largest H\,I absorption signature corresponds with a velocity value of $-$46 km s$^{-1}$ (near to $-$ 41 km s$^{-1}$ given by \citealt{bro96}). The regions of {\rm DBS}\,77 and {\rm DBS}\,78 are relatively near ($\sim 15^{\arcmin}$) in the same place of the Galaxy so for {\rm DBS}\,78 we used the same rotation curves that in the previous case (see Fig.~\ref{Grup300}) therefore we adopted also the same distance value (4.6 kpc).

The corresponding photometric diagrams reveals that {\rm DBS}\,78 is suffering a stronger absorption than {\rm DBS}\,77 (see Table \ref{tab:param}). Our photometric analysis indicated the presence of 15 B type stars and, $\sim$ 20 probable PMS stars for {\rm DBS}\,78. In the region of this cluster there are 2 {\rm GLIMPSE} sources with 3 possible IR counterparts and 1 {\rm MSX} source with 4 possible IR counterparts.

\subsection{{\rm DBS}\,102}
In this case, we analyzed our photometric data centering on two regions:
\begin{itemize}
\item {\rm DBS}\,102a: This region is centered at the coordinates given for \citet{dut03} for the cluster and associated also with {\rm IRAS} source 16112-4943.
\item {\rm DBS}\,102b: This region is centered near the location of {\rm MSX}6C G333.0058+00.7707 and {\rm IRAS} 16115-4941 sources where $JHK$ images revealed the presence of a small H\,II region.
\end{itemize}
Since in these regions there are not available stars with spectral classification, we employed the 21$-$cm H\,I emission of the {\rm DBS}\,102 region to obtain the kinematic distance. We adopted the distance provided for the Galactic kinematic models using \citet{fic89} data. We could notice at Fig. \ref{DBS102RV} that the main H\,I absorption feature in this line of sight is present from $\simeq~-$~52~km~s$^{-1}$ to 0 km s$^{-1}$. The former value is consistent with the associated with the H\,II region G333.0+0.8 given by \citet{kuc97}. In this way it was possible to resolve the ambiguity in distance of the adopted kinematic model and to assign a kinematic distance of 3.6 $\pm$ 0.4 kpc. If the H\,II region would be at the other distance ($\simeq$ 12 kpc) obtained with this rotation model of the Galaxy, H\,I absorption would have to be observed at radial velocity values lower than $-$60 km s$^{-1}$ and this is not the case.

Applying our photometric analysis, we could identify 1 O type star, 9 B type stars and approximately 13 PMS stars in all {\rm DBS}\,102 region. There are 8 GLIMPSE sources with 9 possible IR counterparts. 

 \begin{figure}[!ht]
  \centering
  \includegraphics[width=0.4\textwidth]{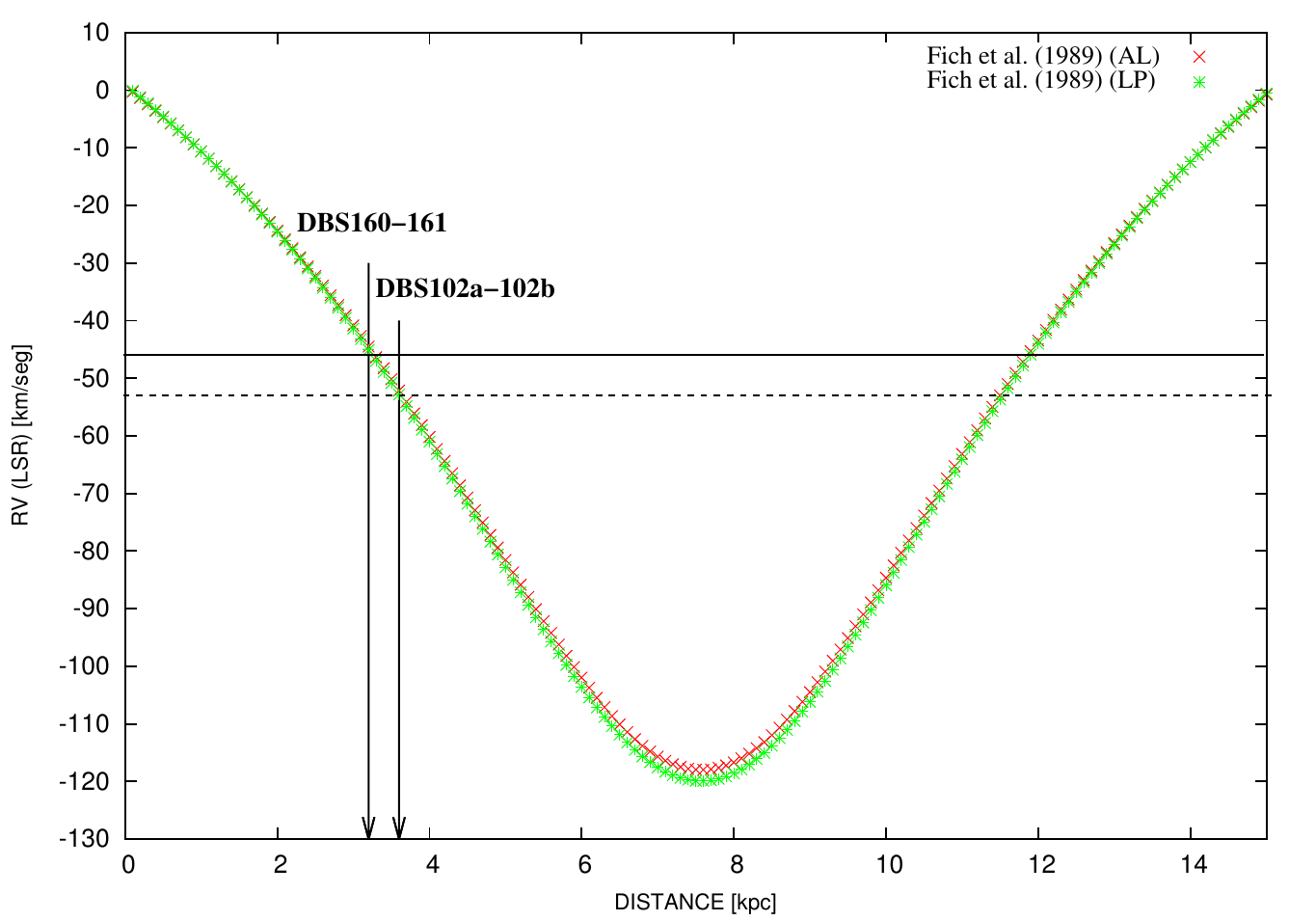} 
  \caption{Galactic rotation models applied at l = 333$^{\circ}$ to our Galaxy. Arrows indicate the corresponding distance to the radial velocity of the radio sources according to the model.}
  \label{Grupo332}
\end{figure}
 \subsection{{\rm DBS}\,~160 and {\rm DBS}\,~161}
Working with the MS stars in the region with spectral classification \citep{rom09}, we obtained an spectrophotometric distance of 2.9 kpc, consistent with the kinematic one, obtained from \citet{fic89} data (3 $\pm$ 0.3 kpc) and close to the spectrophotometric mean value of 2.4 $\pm$ 0.5 kpc obtained by \citet{rom09}. The analysis of the Galactic rotation curve is similar to that performed for {\rm DBS}\,102 region.

Our photometric study revealed 1 O type star, 28 B type stars and approximately 20 PMS stars for both clusters, {\rm DBS}\,~160 and {\rm DBS}\,~161. 
Thus, in all the {\rm DBS}\,~160$-$161 region we found a less number of OB type stars than \citet{rom04}. They indicated the presence of 3 O type stars and 55 B type stars. Probably, this difference is the consequence of the spectral classification performed with photometric analysis. 

\begin{figure}[!ht]
  \centering
 \includegraphics[width=0.5\textwidth]{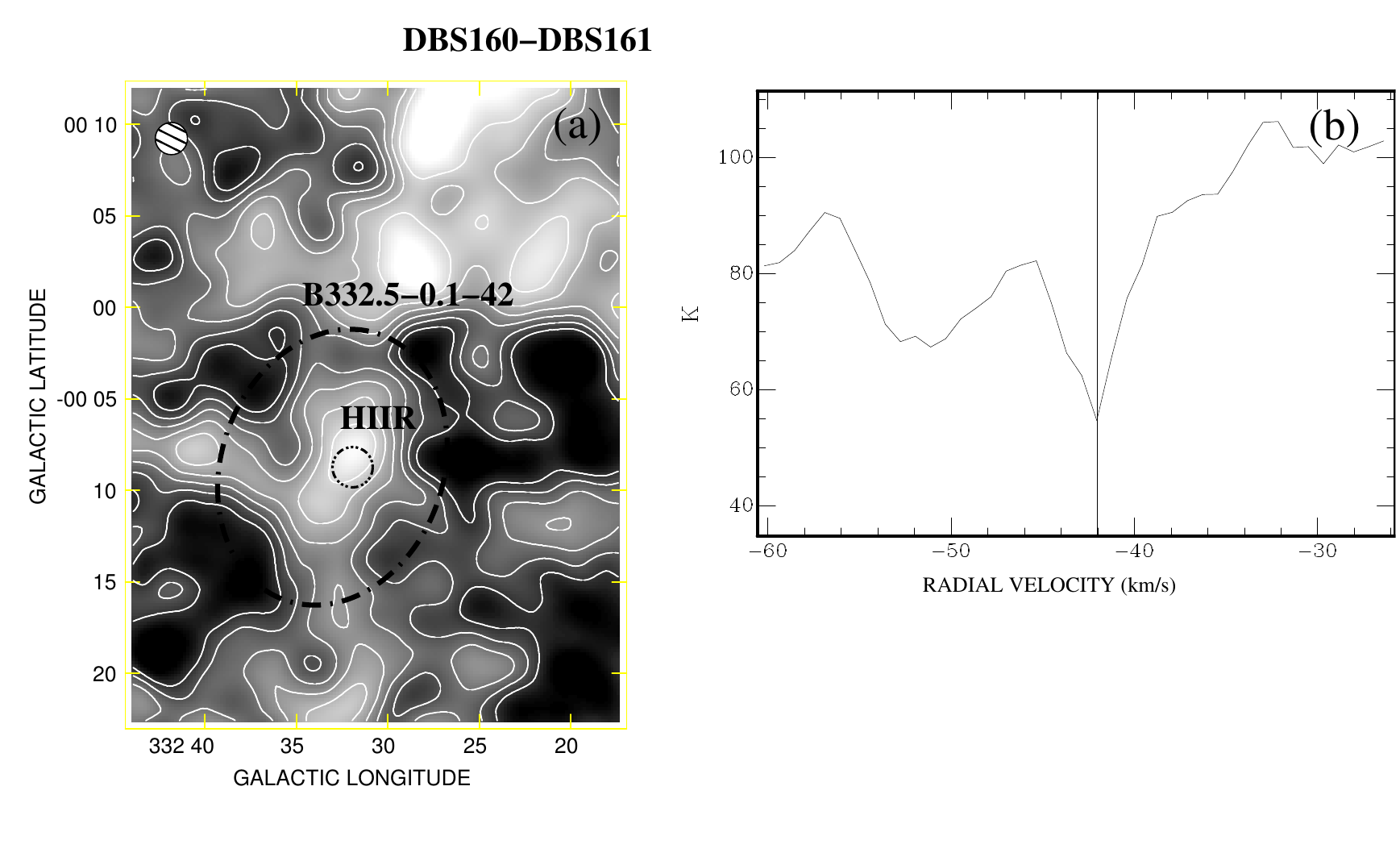}
\caption{{\rm DBS}\,160-161: (a) Mean $T_b$ of the H\,I emission distribution associated with the {\rm B}332.5$-$0.1$-$42 in the velocity range from  $-$44.5 to $-$40.4 km s$^{-1}$. Contour levels are from 80 to 110 K in steps of 5 K. Best fitted ellipse to the bubble is marked with dash lines and the circle inside it indicated the ultra compact H\,II region {\rm GRS}332.5$-$00.10. (b) H\,I brightness temperature profile along $b = -0\fdg14$ where the vertical line indicates the minimum emission corresponding to the systemic velocity.}
\label{fig:bubble}
\end{figure}

The H\,I emission in this region (see Fig. \ref{DBS160RV}a) revealed the presence of a bubble located at $(RV, b) \sim (-$ 42 km s$^{-1}, -$0\fdg14). Figure \ref{fig:bubble}b shows a cross$-$cut of Fig. \ref{DBS160RV}a along $b = -$0\fdg14 where could be observed the presence of two peaks in the $T_{b}$ at $-45.9$ ($v_m$) and $-38$ ($v_M$) km s$^{-1}$ that defines the walls of the bubble. Under the assumption of a symmetric expansion of the bubble,  it will attains its maximum dimension at the systemic velocity ($v_0$). The value of $v_0$  corresponds to the velocity between the two peaks where the minimum value of $T_{b}$ is observed. At intermediate velocities the dimension of the bubble shrinks as $v_M$ (or $v_m$) is approached. The expansion velocity ($v_{exp}$) is estimated to be half of the total velocity range ( $v_{exp} =$ 0.5 $\mid$ $v_M - v_m$ $\mid$) covered by the HI emission related to the feature. 
In Fig. \ref{fig:bubble}a is shown the  HI emission distribution in the velocity range from $-$~44.5 to $-$~40.4 km~s$^{-1}$, where the bubble is easily recognizable. We will refer to this bubble as {\rm B}332.5$-$0.1$-$42.

To estimate the main physical parameters of {\rm B}332.5$-$0.1$-$42, we characterized the ellipse obtained from the least-squares fit to the local maxima around the cavity.
The parameters derived from this fit were the symmetry center of the ellipsoidal HI distribution ($l_0, b_0$), the length of both the semi-major (a) and semi-minor (b) axes of the ellipse, and the inclination angle ($\phi$) between the major axis and the galactic longitude axis (measured positive toward the north Galactic pole). 

We also estimated the total gaseous mass associated with {\rm B}332.5$-$0.1$-$42 as $M_{\rm {H\,I}}[gr] = N_{\rm {H\,I}}[cm^{-2}]\, m_{\rm {H\,I}}[gr]\,A_{\rm {H\,I}}[cm^2]$, where $N_{\rm {H\,I}}$ is the H\,I column density, $N_{\rm {H\,I}} = C\int^{v_m}_{v_M} T_{b} \, dv $, and $A_{\rm {H\,I}}$ is the area covered by the bubble computed from the solid angle covered by the structure at the adopted distance to the Sun. 
Adopting solar abundances, the total gaseous mass of the bubble is M$_t$ = 1.34 M$_{HI}$.

Another important parameter that characterizes the bubble is the kinetic energy, which is given by $E_k [erg] = 0.5\,M_t[gr]\, v^2_{exp}[cm^2\,s^{-2}]$. The dynamic age of the bubble could be estimated from $t_{dyn}[Myr] = \alpha\, R_{ef}[pc]/v_{exp}[km s^{-1}]$, where  $\alpha = 0.6$ for a stellar wind bubble \citep{wea77}.

All the parameters computed for the bubble {\rm B}332.5$-$0.1$-$42 are presented in Table \ref{tab:bubble} in self-explanatory format.

\begin {table}
\caption{Main parameters of the {\rm B}332.5$-$0.1$-$42}
\centering
    \begin{tabular}{c c }
 \hline\hline
  Parameter & Value \\
\hline
(l$_0$,b$_0$) [$^\circ$] & (332.54, -0.14) \\
a [$^\circ$] & 0.14 $\pm$ 0.01 \\
b [$^\circ$] & 0.10 $\pm$ 0.01 \\
$\phi$  [$^\circ$] & 60\\
d [kpc] & 3.1 $\pm$ 0.3 \\
a [pc] & 7.6 $\pm$ 1.1\\
b [pc] &  5.4 $\pm$ 0.8 \\
v$_0$ [km s$^{-1}$] & $-$42.1 $\pm$ 0.8 \\
v$_\mathrm{exp}$ [km s$^{-1}$] & 4.0 $\pm$ 0.8 \\
M$_\mathrm{t}$ [M$_\odot$] & 515 $\pm$ 130    \\
{E}$_\mathrm{k}$ [erg] & ($5.9 \pm 2.8 )\times 10^{46}$\\
t ($\alpha = 0.6$) [Myr] & 1.0 $\pm$ 0.3  \\
\hline
\label{tab:bubble}
 \end{tabular}
\end{table}

\section{Particular discussions}
\subsection{Was the H\,I bubble ({\rm B}332.5$-$0.1$-$42) created by the stars of {\rm DBS}\,160 and {\rm DBS}\,161 clusters?}
\label{sec:discussion}
According with our analysis, it seems that {\rm DBS}\,160$-$161 clusters are located inside the {\rm B}332.5$-$0.1$-$42 bubble. Is it possible that these stars with their winds could have created the bubble?
In order to answer this question we calculated the total mechanical energy ({\it E$_w$}) injected by each star in the course of its entire evolutionary stage. This value was estimated as {\it E$_W$=L$_w$$\tau$}, where $\tau$ is the age of the star and {\it L$_w$} is the stellar mechanical luminosity {\it (L$_w$= 0.5\.Mv$_\infty^2$)} obtained from the stellar mass loss rate {\it ($\dot{M}$)} and the wind terminal velocity {\it ($v_\infty$)}. We estimated these latter two parameters by using the atmosphere models derived by \citet{vin01}. They {\it predict the rate at which mass is lost due to stellar winds from massive O and B-type stars as a function of metal abundance}. We employed the results presented in their Table 3 using solar metallicity and the corresponding $T_{eff}$ for each star assigned by \citet{lando82} from spectral classification.

The stellar mass for each spectral type was obtained according to the analysis developed in Sect. \ref{stellarmass}. In these computations we adopted a model of a single stellar population for the massive cluster stars. Therefore, the estimated age of the cluster could be considered as the same value ($\tau$) for all the cluster brightest members ({\rm MS} stars). We obtained in this way an upper limit for the total mechanical energy injected by the cluster stars into the {\rm ISM}. To avoid field stellar contamination to the cluster stars, we applied the above procedure to compute the total energy to the stars located in the clusters region and to the stars in the adopted comparison field, and finally we subtracted both results. Thus the net total mechanical energy was (1.43  $\pm$ 0.52) $\times$ 10$^{50}$ erg.

The theoretical models predict that 20\% only of the wind-injected energy is converted into the mechanical energy of an expanding shell \citep{wea77}.
The energy conversion efficiency seems to be as low as 2--5 \% \citep{cap03} from an observational viewpoint. In this way, the energy used to originate the bubble may have been (2.9 or 7.1) $\times$ 10$^{48}$ erg.

Both these values are greater than the kinetic energy that would actually have the {\rm G}332.5$-$0.1$-$42 bubble. 
With this result, the stars possible members of {\rm DBS}\,160 and {\rm DBS}\,161 clusters could have generated the bubble.

\subsection{The evolutionary state of the studied H\,II regions}
\label{sec:HIIregions}
All embedded clusters analyzed in this paper share their location in the Galaxy with H\,II regions. For {\rm DBS}\,77, {\rm DBS}\,102ab and {\rm DBS}\,160$-$161 clusters, we obtained that {\rm U} $>$ {\it u}, so, the neutral gas cloud where the H\,II region is immersed is completely ionized, therefore {\it the region is limited by density}. Beyond this material, there is not ionizable material or there is dust mixed with the ionized gas that absorb UV photons emitted by the cluster stars and then re-emits them in the IR wavelength. The photons are consumed and are not useful to ionize the gas, resulting only in an increase of temperature of interstellar dust. This result proves that the short formation phase during which the central cluster rapidly ionizes the neutral medium is over. Therefore, all the studied H\,II regions are in a long expansion phase, where the pressure in the warm ionized gas ($T_e \sim 10^4$ K, see Table \ref{tab:HIIparam}) is higher than $T_b$ in the cold neutral medium surrounding the H\,II regions ($T_b$ $\leq $ 150 K). By this way, the electron density of the H\,II regions will decrease. 

For {\rm DBS}\,78 we found that {\rm U} $<$ {\it u},  therefore {\it the region is limited by ionization}. That is, part of the gas that surrounding the embedded cluster 
has been ionized by exciters stars and the other part of the gas has not been ionized yet. This may be due to the absence in the cluster of O type stars, or by not considering the Lyman photons emission produced by the PMS stars present in {\rm DBS}\,78. 

The infrared emission from dust around developed H\,II regions argue strongly that most H\,II regions do form inside dusty H\,I regions. A small H\,II region develops around the star, probably in material rather denser than average for the cloud, due to the local concentration which produced the star initially.

\section{Conclusions}
\label{sec:conclusions}
We studied the embedded clusters {\rm DBS}\,77, 78, 102, 160, and 161 placed behind dust clouds in the fourth quadrant of the V\'{\i}a L\'actea. We grouped these clusters into four regions, according to their spatial location and we computed their fundamental parameters, revealing the nature of their most brightest members and the probable presence of PMS stars in all of them (see Tables \ref{tab:param} and \ref{tab:sptphot} for details).

We found that the {\rm DBS}\,160$-$161 clusters are located inside the H\,I bubble designed by us as {\rm B}332.5$-$0.1$-$42 and we estimated the main parameters of this bubble (see Table \ref{tab:bubble} for details). The stellar winds of {\rm DBS}\,160$-$161 clusters members may well have been the creators of {\rm B}332.5$-$0.1$-$42.

We also obtained the IMF for massive stars members of each cluster. The slope value obtained for {\rm DBS}\,77 cluster ($\Gamma = -$1.44) was similar to $\Gamma = -$1.35 given by \citet{sal55}. The slope values obtained for {\rm DBS}\,160$-$161 clusters ($\Gamma = -$0.71), for {\rm DBS}\,102ab ($\Gamma = -$0.75) and for {\rm DBS}\,78 ($\Gamma = -$1.19), were a little less steeper than \citet{sal55} value. This behavior was also found in some young stellar clusters \citep[see][]{bau03} and could be a signature in this kind of objects.

In addition to the clusters mentioned in the previous paragraph, we studied the H\,II regions associated with each of them. We estimated the main parameters of these H\,II regions (see Table \ref{tab:HIIparam}). We computed their mean {\rm LSR} radial velocity (RV$_{(LSR)}$) to obtain a kinematic distance and compare their values with the spectrophotometric ones. Regarding the angular diameters, $\theta_{sph}$, we could identify the H\,II region in the {\rm DBS}\,102 cluster as an extended one.
We could estimate the excitation parameter {\it u} of each H\,II region and the ionization parameter {\it U} of each cluster linked to these H\,II regions. The comparison of {\it u} with {\it U} is a basic tool in the study of relations between H\,II regions and exciting stars. We obtained that the regions linked to the {\rm DBS}\,77, {\rm DBS}\,102ab and {\rm DBS}\,160$-$161 embedded clusters, are limited by density. This result indicates that the central cluster has already ionized all the interstellar material and the studied H\,II regions are in a long expansion phase, where the pressure in the warm ionized gas is higher than the same in the cold neutral medium surrounding the H\,II regions. The region linked to {\rm DBS}\,78 embedded cluster is limited by ionization. This result indicates that at the moment there has been ionized only a part of the region.

\begin{acknowledgements}
MAC, GLB, JAP, LAS y JCT acknowledges support from 
CONICET (PIPs 112-201201-00226 y 112-201101-00301).
The authors are much obliged for the use of the NASA Astrophysics Data System,
of the $SIMBAD$ database and $ALADIN$ tools (Centre de Donn\'es Stellaires
--- Strasbourg, France).
This publication also made use of data from:
a) the Two Micron All Sky Survey, which is a joint project of the University
of Massachusetts and the Infrared Processing and Analysis Center/California
Institute of Technology, funded by the National Aeronautics and Space
Administration and the National Science Foundation;
b) the AAVSO Photometric All-Sky Survey (APASS), funded by the Robert Martin
Ayers Sciences Fund.
c) the Midcourse Space Experiment ({\rm MSX}). Processing of the data was funded by
the Ballistic Missile Defense Organization with additional support from NASA
Office of Space Science.
Support for JB, SRA and RK is provided by the Ministry of Economy, Development, and Tourism's Millennium Science Initiative through grant IN 120009,  awarded to The Millennium Institute of Astrophysics, MAS. JB is supported by FONDECYT No.1120601, RK is supported  by Fondecyt Reg. No. 1130140, SRA by Fondecyt No. 3140605.
We thank R. Mart\'{\i}nez and H. Viturro for technical support and E. Marcelo Arnal for many useful comments which improved this paper. Finally, we wish to thank the anonymous referee for the suggestions and comments that improved the original version of this work.

\end{acknowledgements}

\newpage

\begin{table*}
\scriptsize
\caption{Adopted photometric data and spectral classification for the most relevant stars on each studied region}
\centering
\begin{tabular}{rccrrrrrrccrrrr}
\hline
\hline
$ID$ & $\alpha_{J2000}$ & $\delta_{J2000}$ & $B$ & $V$ & $I$ & $J$  & $H$ & $K$ & $ST$ & S/N & $E_{B-V}$ & $E_{J-K}$ & $V_0-M_V$ \\
     & $[h:m:s]$ & $[^{\circ}:^{\arcmin}:^{\arcsec}]$ & $[mag]$ & $[mag]$ & $[mag]$ & $[mag]$ & $[mag]$ & $[mag]$ & &     & $[mag]$ & $[mag]$ & $[mag]$   \\
\hline
{\rm DBS}\,77 $+$ {\rm VVV}\,16 & & & & & & & & & & & & &   \\
\hline
   3\tablefootmark{$\bullet$} & 12:34:57.1 & -61:37:23.6 &  $--$ &  $--$ &  $--$ &  7.74 &  6.63 &  6.21 & K0$-$2\,III &   69 & $--$ & $--$ & $--$ \\
   5\tablefootmark{$\bullet$} & 12:34:41.3 & -61:39:23.5 & 14.64 & 13.12 & 10.05 &  7.90 &  7.54 &  6.93 & K5\,III   &   84 & 0.00 & $--$ & 13.3 \\
   8\tablefootmark{$\bullet$} & 12:34:49.5 & -61:38:20.2 & 12.95 & 11.49 & 10.36 &  8.97 &  8.24 &  8.08 & K0$-$2\,III &   80 & 0.38 & $--$ &  9.7 \\
  18~~           & 12:34:50.7 & -61:39:27.1 & 13.62 & 12.20 & 10.70 &  9.70 &  9.07 &  8.90 & O9-B1I  &  106 & 1.65 & $--$ & 13.5 \\
  21~~           & 12:35:00.4 & -61:40:22.9 & 14.45 & 12.99 & 11.31 & 10.20 &  9.58 &  9.42 & O7-8V   &   86 & 1.78 & 1.00 & 12.7 \\
  27\tablefootmark{$\bullet$} & 12:34:51.4 & -61:38:40.7 & 15.75 & 14.12 & 11.86 & 10.44 &  9.93 &  9.65 & K0III   &   83 & 0.63 & $--$ & 11.5 \\
  39~~           & 12:35:00.1 & -61:41:36.8 & 15.64 & 14.13 & 12.09 & 10.83 & 10.45 & 10.20 & O8V     &   95 & 1.83 & 0.85 & 13.5 \\
  62~~           & 12:34:58.9 & -61:39:48.2 & 20.82 & 18.18 & 14.69 & 12.23 & 11.24 & 10.72 & o8.4v   & $--$ & 2.94 & 1.70 & 13.2 \\
  87~~           & 12:35:02.0 & -61:42:03.5 & 17.34 & 15.52 & 13.48 & 12.10 & 11.42 & 11.10 & b0.2v   & $--$ & 2.13 & 1.16 & 13.2 \\
 103\tablefootmark{$\bullet$} & 12:35:00.0 & -61:39:02.0 & 16.80 & 15.17 & 13.32 & 12.07 & 11.37 & 11.27 & G4III   &   87 & 0.79 & $--$ & 11.8 \\
 128~~           & 12:34:52.6 & -61:40:19.2 & 21.17 & 19.39 & 15.60 & 13.11 & 12.17 & 11.66 & b0.6v   & $--$ & $--$  & 1.60 & 13.2 \\
 132~~           & 12:35:00.7 & -61:41:16.2 & 18.81 & 17.01 & 14.62 & 12.96 & 12.14 & 11.69 & b1.1v   & $--$ & 2.04 & 1.41 & 13.2 \\
 144~~           & 12:35:01.2 & -61:41:32.9 & 19.64 & 18.04 & 15.67 & 13.88 & 12.82 & 11.89 & pms     & $--$ & $--$ & $--$ & 13.2 \\
 203~~           & 12:34:50.7 & -61:39:52.3 &  $--$ &  $--$ &  $--$ & 14.10 & 13.00 & 12.42 & b1.4v   & $--$ & $--$ & 1.82 & 13.2 \\
 240~~           & 12:35:03.6 & -61:41:45.4 & 24.97 & 21.57 & 17.63 & 15.33 & 13.70 & 12.69 & b1v     & $--$ & 3.65 & 2.78 & 13.2 \\
 249~~           & 12:34:51.2 & -61:40:15.8 &  $--$ &  $--$ &  $--$ & 14.10 & 13.22 & 12.73 & b2.2v   & $--$ & $--$ & 1.49 & 13.2 \\
 258~~           & 12:34:50.2 & -61:39:21.0 &  $--$ &  $--$ &  $--$ & 16.37 & 14.20 & 12.79 & pms     & $--$ & $--$ & $--$ & 13.2 \\
\hline
\multicolumn{14}{l}{DBS 78} \\
\hline
  78 & 12:36:05.6 & -61:50:46.8 & 20.29 & 19.42 & 14.87 & 12.88 & 11.90 & 10.95 & pms   & $--$ & $--$ & $--$ & 13.2 \\
 100 & 12:36:02.7 & -61:50:44.6 &  $--$ &  $--$ &  $--$ & 15.02 & 12.77 & 11.26 & pms   & $--$ & $--$ & $--$ & 13.2 \\
 113 & 12:36:00.2 & -61:51:10.9 &  $--$ &  $--$ &  $--$ & 13.53 & 12.47 & 11.50 & pms   & $--$ & $--$ & $--$ & 13.2 \\
 229 & 12:36:05.4 & -61:50:44.9 &  $--$ &  $--$ &  $--$ & 16.46 & 14.18 & 12.55 & pms   & $--$ & $--$ & $--$ & 13.2 \\
 253 & 12:35:57.7 & -61:51:12.0 &  $--$ &  $--$ &  $--$ & 16.68 & 14.47 & 12.75 & pms   & $--$ & $--$ & $--$ & 13.2 \\
 330 & 12:36:04.8 & -61:51:02.0 &  $--$ &  $--$ &  $--$ & 16.93 & 14.66 & 13.02 & pms   & $--$ & $--$ & $--$ & 13.2 \\
 347 & 12:36:01.4 & -61:50:53.5 &  $--$ &  $--$ &  $--$ & 15.72 & 14.19 & 13.08 & pms   & $--$ & $--$ & $--$ & 13.2 \\
 386 & 12:36:03.3 & -61:50:56.4 &  $--$ &  $--$ &  $--$ & 16.51 & 14.59 & 13.21 & pms   & $--$ & $--$ & $--$ & 13.2 \\
 416 & 12:36:04.7 & -61:50:55.7 &  $--$ &  $--$ &  $--$ & 16.29 & 14.40 & 13.32 & b1.4v & $--$ & $--$ & 3.11 & 13.2 \\
 466 & 12:36:02.2 & -61:50:53.0 &  $--$ &  $--$ &  $--$ & 15.87 & 14.33 & 13.43 & b2.2v & $--$ & $--$ & 2.57 & 13.2 \\
 495 & 12:36:08.6 & -61:51:06.4 &  $--$ &  $--$ &  $--$ & 16.00 & 14.41 & 13.51 & b2.2v & $--$ & $--$ & 2.62 & 13.2 \\
 510 & 12:36:04.8 & -61:51:11.5 &  $--$ &  $--$ &  $--$ & 16.40 & 14.56 & 13.54 & b1.8v & $--$ & $--$ & 2.99 & 13.2 \\
 520 & 12:36:02.4 & -61:51:03.4 &  $--$ &  $--$ &  $--$ & 15.59 & 14.26 & 13.57 & b2.6v & $--$ & $--$ & 2.13 & 13.2 \\
 623 & 12:36:01.3 & -61:51:17.4 &  $--$ &  $--$ &  $--$ & 16.35 & 14.93 & 13.80 & pms   & $--$ & $--$ & $--$ & 13.2 \\
 654 & 12:36:02.2 & -61:52:06.3 &  $--$ &  $--$ &  $--$ & 16.35 & 14.87 & 13.87 & pms   & $--$ & $--$ & $--$ & 13.2 \\
 692 & 12:36:02.3 & -61:50:41.4 &  $--$ &  $--$ &  $--$ & 17.19 & 15.26 & 13.95 & pms   & $--$ & $--$ & $--$ & 13.2 \\
 699 & 12:36:00.7 & -61:50:29.6 &  $--$ &  $--$ &  $--$ & 18.15 & 15.57 & 13.96 & pms   & $--$ & $--$ & $--$ & 13.2 \\
\hline
\multicolumn{14}{l}{DBS 102a + DBS 102b} \\
\hline
  20 & 16:15:02.0 & -49:50:40.1 &  $--$ &  $--$ & $--$ & 10.92 &  9.85 &  9.28 & o4.9v & $--$ & $--$ & 1.86 & 12.6 \\
  92 & 16:15:18.7 & -49:48:53.5 &  $--$ &  $--$ & $--$ & 13.50 & 11.80 & 10.61 & pms   & $--$ & $--$ & $--$ & 12.6 \\
 175 & 16:15:12.7 & -49:49:11.8 &  $--$ &  $--$ & $--$ & 12.99 & 12.22 & 11.64 & pms   & $--$ & $--$ & $--$ & 12.6 \\
 183 & 16:15:12.9 & -49:49:01.0 &  $--$ &  $--$ & $--$ & 13.69 & 12.56 & 11.74 & pms   & $--$ & $--$ & $--$ & 12.6 \\
 252 & 16:15:00.3 & -49:50:48.6 &  $--$ &  $--$ & $--$ & 13.40 & 12.61 & 12.22 & b2.4v & $--$ & $--$ & 1.29 & 12.6 \\
 280 & 16:15:01.1 & -49:50:42.5 & 20.60 & 18.90 & $--$ & 14.50 & 13.09 & 12.36 & b1.6v & $--$ & 1.68 & 2.28 & 12.6 \\
 283 & 16:14:58.6 & -49:50:09.0 & 21.81 & 21.03 & $--$ & 16.09 & 13.90 & 12.36 & pms   & $--$ & $--$ & $--$ & 12.6 \\
 301 & 16:15:03.1 & -49:50:04.6 &  $--$ &  $--$ & $--$ & 16.00 & 13.85 & 12.44 & pms   & $--$ & $--$ & $--$ & 12.6 \\
 426 & 16:15:00.8 & -49:50:27.8 & 18.86 & 17.87 & $--$ & 16.68 & 14.56 & 12.87 & pms   & $--$ & $--$ & $--$ & 12.6 \\
 468 & 16:15:13.7 & -49:49:14.5 &  $--$ &  $--$ & $--$ & 15.78 & 13.96 & 12.96 & b1.8v & $--$ & $--$ & 2.94 & 12.6 \\
 476 & 16:15:14.8 & -49:48:10.3 &  $--$ &  $--$ & $--$ & 17.50 & 14.85 & 12.99 & pms   & $--$ & $--$ & $--$ & 12.6 \\
 608 & 16:15:01.4 & -49:50:50.5 &  $--$ &  $--$ & $--$ & 16.32 & 14.45 & 13.33 & b2.4v & $--$ & $--$ & 3.11 & 12.6 \\
 777 & 16:14:59.5 & -49:50:35.0 & 24.56 & 21.86 & $--$ & 16.93 & 14.99 & 13.69 & pms   & $--$ & $--$ & $--$ & 12.6 \\
 808 & 16:15:00.4 & -49:50:01.3 & 25.49 & 22.79 & $--$ & 16.49 & 14.77 & 13.76 & b4.4v & $--$ & 2.38 & 2.82 & 12.6 \\
 876 & 16:15:00.5 & -49:50:47.2 &  $--$ &  $--$ & $--$ & 16.78 & 14.95 & 13.88 & b4.4v & $--$ & $--$ & 2.98 & 12.6 \\
\hline
\multicolumn{14}{l}{DBS 160 + DBS 161} \\
\hline
  77 & 16:17:02.2 & -50:47:03.2 & $--$ & $--$ & $--$ & 11.97 & 10.65 &  9.73 & pms$^{(1)}$   & $--$ & $--$ & $--$ & 12.3 \\
  93 & 16:17:05.0 & -50:47:25.7 & $--$ & $--$ & $--$ & 11.20 & 10.37 &  9.94 & b0.1v     & $--$ & $--$ & 1.44 & 12.3 \\
 122 & 16:17:09.2 & -50:47:14.7 & $--$ & $--$ & $--$ & 13.26 & 11.47 & 10.15 & pms       & $--$ & $--$ & $--$ & 12.3 \\
 126 & 16:16:55.7 & -50:47:23.0 & $--$ & $--$ & $--$ & 11.65 & 10.79 & 10.21 & pms$^{(2)}$   & $--$ & $--$ & 1.62 & 12.5 \\
 264 & 16:17:06.0 & -50:46:56.0 & $--$ & $--$ & $--$ & 15.33 & 13.19 & 11.07 & pms       & $--$ & $--$ & $--$ & 12.3 \\
 382 & 16:16:50.5 & -50:47:44.9 & $--$ & $--$ & $--$ & 11.98 & 11.73 & 11.73 & YSO$^{(3)}$   & $--$ & $--$ & $--$ & 12.3 \\
 437 & 16:17:06.3 & -50:47:09.3 & $--$ & $--$ & $--$ & 13.39 & 12.42 & 11.92 & b2.1v     & $--$ & $--$ & 1.59 & 12.3 \\
 450 & 16:17:02.6 & -50:46:56.7 & $--$ & $--$ & $--$ & 14.54 & 12.94 & 11.97 & b1.4v     & $--$ & $--$ & 2.70 & 12.3 \\
 555 & 16:17:04.6 & -50:47:25.7 & $--$ & $--$ & $--$ & 14.83 & 13.20 & 12.29 & b1.7v     & $--$ & $--$ & 2.66 & 12.3 \\
 642 & 16:17:08.6 & -50:47:11.6 & $--$ & $--$ & $--$ & 15.30 & 13.61 & 12.50 & pms       & $--$ & $--$ & $--$ & 12.3 \\
 662 & 16:16:50.9 & -50:47:44.0 & $--$ & $--$ & $--$ & 12.76 & 12.50 & 12.53 & B8-9V$^{(3)}$ & $--$ & $--$ & 0.26 & 12.2 \\
1005 & 16:17:10.5 & -50:47:11.7 & $--$ & $--$ & $--$ & 15.29 & 13.89 & 13.10 & b4.4v     & $--$ & $--$ & 2.27 & 12.3 \\
1064 & 16:16:56.0 & -50:47:20.6 & $--$ & $--$ & $--$ & 14.22 & 13.56 & 13.16 & b8.2v     & $--$ & $--$ & 1.10 & 12.3 \\
1206 & 16:16:55.6 & -50:47:26.9 & $--$ & $--$ & $--$ & 14.90 & 13.87 & 13.33 & b7.2v     & $--$ & $--$ & 1.61 & 12.3 \\
1216 & 16:16:58.4 & -50:47:32.8 & $--$ & $--$ & $--$ & 14.56 & 13.79 & 13.34 & b8.4v     & $--$ & $--$ & 1.25 & 12.3 \\
1246 & 16:17:04.1 & -50:47:44.5 & $--$ & $--$ & $--$ & 14.93 & 13.90 & 13.38 & b7.3v     & $--$ & $--$ & 1.60 & 12.3 \\
1275 & 16:17:01.8 & -50:47:26.5 & $--$ & $--$ & $--$ & 14.47 & 13.78 & 13.40 & b8.8v     & $--$ & $--$ & 1.09 & 12.3 \\
1340 & 16:16:55.3 & -50:47:25.6 & $--$ & $--$ & $--$ & 16.03 & 14.43 & 13.47 & b5.4v     & $--$ & $--$ & 2.64 & 12.3 \\
\hline
\label{tab:sptphot}
\end{tabular}
\tablefoot{
$V_0-M_V$ are the individual values for stars with spectral data, but the adopted cluster values (see Table \ref{tab:param}) for stars with only photometric information.
\tablefoottext{$^{\bullet}$}{Indicates adopted no cluster member stars.}
\tablefoottext{1}{SpT = o5v asigned by \citet{rom04}$-$IRS1; not considered to calculate the mechanical energy (see subsection \ref{sec:discussion}).}
\tablefoottext{2}{Considered O9.5 V spectroscopic spectral type according to \citet{rom09}$-$IRS3 to calculate the mechanical energy (see subsection \ref{sec:discussion}).}
\tablefoottext{3}{SpT asigned by \citet{rom04}.}
}
\end{table*}

\bibliographystyle{aa}  
\bibliography{bibliojulio2015}

\end{document}